\documentclass[11pt]{article}
\usepackage[active]{srcltx}
\usepackage[T1]{fontenc}
\usepackage[utf8]{inputenc}
\usepackage{lmodern}
\usepackage[a4paper]{geometry}
\usepackage{amsmath,amssymb,amsthm,mathabx}
\usepackage{pgf,tikz}
\usepackage{mathrsfs}
\usepackage{graphicx}
\usepackage{subcaption}
\usepackage{mwe}
\usepackage[colorlinks=true,breaklinks=true,linkcolor=blue,urlcolor=green,citecolor=red]{hyperref} 
\usepackage{yfonts}
\usepackage{cancel}
\usetikzlibrary{arrows}

\renewcommand{\epsilon}{\varepsilon}

\theoremstyle{remark}

\usepackage{bm}
\usepackage{enumitem}

\usepackage{appendix}

\newcounter{movie}

\renewcommand{\themovie}{M\arabic{movie}}

\newenvironment{movie}[1]
{%
    \refstepcounter{movie}%
    \par\medskip
    \noindent\textbf{Movie~\themovie: }\label{movie:#1}%
}
{\par\medskip}

\newcommand{\movieref}[1]{\ref{#1}}

\usepackage{soul}

\title{Emergent Spatiotemporal Dynamics in Large-Scale Brain Networks with Next Generation Neural Mass Models}
\date{}
\author{Rosa Maria Delicado$^{1}$, Gemma Huguet$^{2,3,4}$, Pau Clusella$^{2}$ \\
\parbox{12.5cm}{
  \small
  \begin{itemize}
    \item[$^1$] Departament de Matem\`atiques i Inform\`atica and Institute of Applied Computing and Community Code (IAC3), Universitat de les Illes Balears, Spain
  \item[$^2$]
    Departament de Matem\`atiques, Universitat Polit\`ecnica de Catalunya, Barcelona, Spain 
  \item[$^3$]
    Institut de Matem\`atiques de la UPC - Barcelona Tech (IMTech), Barcelona, Spain 
  \item[$^4$]
  Centre de Recerca Matem\`atica, Barcelona, Spain
   \end{itemize}
 }}
 \date{}

\begin{document}

\maketitle

\textbf{Corresponding author:} \texttt{rosa-maria.delicado@uib.cat} \\


\abstract{Understanding the dynamics of large-scale brain models remains a central challenge due to the inherent complexity of these systems.
In this work, we explore the emergence of complex spatiotemporal patterns in a large scale-brain model composed of $90$ interconnected brain regions coupled through empirically derived anatomical connectivity. An important aspect of our formulation is that the  local dynamics of each brain region are described by a next-generation neural mass model, which explicitly captures the macroscopic gamma activity of coupled excitatory and inhibitory neural populations (PING mechanism). 
We first identify the system’s homogeneous states—both resting and oscillatory—and analyze their stability under uniform perturbations.
Then, we determine the stability against non-uniform perturbations by obtaining dispersion relations for the perturbation growth rate. This analysis enables us to link unstable directions of the homogeneous solutions to the emergence of rich spatiotemporal patterns, that we characterize by means of Lyapunov exponents and frequency spectrum analysis.
Our results show that, compared to previous studies with classical neural mass models, next-generation neural mass models provide a broader dynamical repertoire, both within homogeneous states and in the heterogeneous regime. Additionally, we identify a key role for anatomical connectivity in  cross-frequency coupling, allowing for the emergence of gamma oscillations with amplitude modulated by slower rhythms. These findings suggest that such models are not only more biophysically grounded but also particularly well-suited to capture the full complexity of large-scale brain dynamics. Overall, our study advances the analytical understanding of emerging spatiotemporal patterns in whole-brain models.}

\section{Introduction}

Oscillatory and rhythmic activity is ubiquitous in the brain, spanning a wide range of frequency bands—from slow delta (1–4 Hz) and theta (4–8 Hz) rhythms to faster alpha (8-13 Hz), beta (13–30 Hz), and gamma (30–100 Hz) oscillations~\cite{buzsaki2006rhythms}. 
A key challenge in computational neuroscience is to understand how these dynamics emerge and interact across spatial and temporal scales, giving rise to complex spatiotemporal patterns observed in whole brain recordings \cite{muller_cortical_2018,Zhang2018,Halgren2019,deco_turbulent-like_2020}.
Neural mass models (NMMs) represent a primary modeling approach, providing low-dimensional descriptions of population-level dynamics by coarse-graining the activity of large ensembles of neurons and synapses. Canonical formulations such as the Wilson-Cowan \cite{wilson_excitatory_1972}, the Jansen–Rit \cite{jansen_electroencephalogram_1995}, and the Wendling models (reviewed in \cite{Wendling2016}) have been especially successful in reproducing and fitting neuroimaging data, notably EEG and MEG, thereby establishing a principled link between microscopic neural mechanisms and macroscopic brain signals.

By coupling several NMMs using network topologies derived from structural connectomics data, one can produce large-scale brain models (also known as whole-brain models) in which single brain regions described by an NMM interact across nodes. Such models have been widely used to investigate both healthy brain function \cite{muller_cortical_2018,deco_turbulent-like_2020,kunze_transcranial_2016} and a range of pathological conditions, including Parkinson, epilepsy, schizophrenia or Alzheimer's disease \cite{cabral2013structural, stefanovski_linking_2019, saenger2017uncovering,richardson_large_2012, Olmi2019}.
However, these models pose significant challenges for mathematical analysis due to their high dimensionality and the presence of features such as noise, heterogeneities, and transmission delays. 

In \cite{clusella2023complex}, the authors present a simple framework to account for the emergence of rich dynamical phenomena, including high-dimensional chaos and traveling waves. By assuming normalized total inputs received by the nodes, they demonstrate that such complex behaviors can arise purely from network coupling. Their analysis builds on the study of transverse instabilities of homogeneous oscillatory solutions: using the Master Stability Function (MSF) formalism \cite{pecora_master_1998}, they showed how an oscillatory state of identical node dynamics ---modeled by the Jansen–Rit system \cite{jansen_electroencephalogram_1995}--- can lose stability, giving rise to a diverse repertoire of spatiotemporal patterns.

While most large-scale brain models rely on classical neural mass formulations such as Jansen–Rit, recent advances in mean field theory have led to a new generation of neural mass models that provide an exact description of the macroscopic activity of spiking neuronal networks \cite{montbrio2015macroscopic, coombes2019next}. Unlike classical NMMs, the so-called next-generation NMMs (NG-NMMs) capture the exact evolution of the mean firing rate and membrane potential of a population of quadratic integrate-and-fire (QIF) neurons, with dynamics that depend on neuronal synchrony within the population, yielding a richer repertoire of behaviors.  Their combination of accuracy and interpretability has drawn significant attention \cite{dumont2019macroscopic,coombes2019next,bick2020,clusella2022,Coombes2023}, and they are now widely used to address diverse questions in neuroscience \cite{Segneri2020,Taher2020,reyner2022phase,Gerster2021}. 
In particular, incorporating NG-NMM into large-scale brain modeling frameworks represents a promising and rapidly emerging area of research \cite{Gerster2021, Perl2023, Forrester2024}.

Following this line of research, in the present work we investigate the dynamics of a large-scale brain model consisting of a network of 90 nodes, with structural connectivity derived from tractography data \cite{deco2017single}. The activity of each node is governed by a NG-NMM composed of an excitatory and an inhibitory population interacting through exponentially decaying synapses. For this exploration, we adopt parameter values from previous studies \cite{dumont2019macroscopic, reyner2022phase}, thereby representing an instance of the Pyramidal-Interneuron Network Gamma (PING) mechanism for generation of gamma oscillations~\cite{Whittington1995, Whittington2000, Borgers2008, bartos2007,Buzski2012}. 

We provide a thorough analysis of the resulting network dynamics, emphasizing how the incorporation of NG-NMMs shapes large-scale brain activity patterns. Following the approach of \cite{clusella2023complex}, we highlight that such models are capable of generating rich dynamical phenomena, including traveling waves and high-dimensional chaos, under minimal assumptions. 
Our analysis begins by identifying a homogeneous manifold of the system (corresponding to identical node dynamics) that serves as a reference structure for understanding the emergence of more complex spatiotemporal behaviors. 
The dynamics on the manifold are captured by a low-dimensional nonlinear system, which we analyze using numerical continuation methods.
 We then investigate the stability of homogeneous fixed points and limit-cycles to arbitrary perturbations by
decomposing them on a suitable basis, ultimately providing a dispersion relation or Master Stability Function that links
spatial modes with perturbation growth rates.
Simulations and analysis of Lyapunov exponents and frequency spectrum reveal that when this manifold loses stability, it gives rise to a diverse range of dynamical patterns. 
Our results show that this large-scale brain model exhibits a broader repertoire of dynamical behaviors than those reported in \cite{clusella2023complex}, both within the homogeneous manifold and in the heterogeneous regime. Indeed, we find instabilities arising from both stationary and oscillatory states. Among the relevant patterns observed, we highlight the presence of fast oscillations whose amplitude is modulated by a slower rhythm—a pattern that cannot be generated in the isolated nodes. This phenomenon is a type of cross-frequency coupling (CFC) \cite{Buzski2012} widely observed in the brain, particularly in the hippocampus, where theta–gamma coupling structures neural firing into temporally ordered assemblies believed to underlie memory formation and spatial coding \cite{Lisman2013}. Our results suggest that NG-NMMs are not only more biophysically grounded, but also better suited for capturing the full complexity of large-scale brain dynamics.

The paper is organized as follows. In Section \ref{sec:model} we present the mathematical model. In Section \ref{sec:hom} we study the stability of homogeneous states within the homogeneous invariant manifold.
In Section \ref{sec:transverse} we analyze the stability of the homogeneous states with respect to transverse perturbations.  In Section \ref{sec:carac}, we characterize the heterogeneous spatiotemporal patterns that arise when synchronization is lost by means of computations of Lyapunov exponents and frequency spectrum analysis. Finally, we end with a Discussion in Section \ref{sec:discussion}. The Appendix contains details of the linear stability analysis and MSF formalism.

\section{Large-scale brain model}\label{sec:model}

We consider a large-scale brain model consisting of a network with $N=90$ nodes, each representing a distinct brain region, coupled using structural connectivity data obtained in~\cite{deco2017single} from diffusion tensor imaging. The interactions are defined by a $N\times N$ row-normalized connectivity matrix $\tilde{\bm{W}}=(\tilde w_{jk})$, obtained by averaging the connectomes of 16 human subjects and then applying row normalization so that the entries of each row sum to one (as it is usually employed in large scale brain models \cite{Hlinka2012,kunze_transcranial_2016,Roberts2019,Forrester2024}), that is,
\begin{equation}
\label{eq:HomogeneousCondition}
    \displaystyle \sum_{k=1}^N \tilde{w}_{jk} = 1.
\end{equation}
The connection weights $\tilde{w}_{jk}$ are non-negative and indicate the average number of connections from region $k$ to region $j$, for $j,k \in \{ 1,\ldots,N\}$. The network topology is available at \url{github.com/pclus/transverse-instabilities}.

We assume each region is capable of generating oscillatory activity through the local interaction of excitatory and inhibitory neuronal populations, a mechanism that has been associated to the emergence of brain rhythms in the gamma frequency range \cite{Buzski2012}. 
Therefore, the neural activity of each node is modeled using an adapted form of the next-generation neural mass model presented in \cite{dumont2019macroscopic,reyner2022phase}.
Based on an exact low-dimensional description \cite{montbrio2015macroscopic},
this model provides the evolution of the mean firing rate $r$ and mean membrane potential $v$ of an excitatory and an inhibitory population of Quadratic Integrate-and-Fire (QIF) neurons coupled through exponentially decaying synaptic activity $s$. Thus, at each node $j$, the dynamics are governed by the following $6$-dimensional system of differential equations,
\begin{align}
\text{($E_j$)}&\quad\left\{
\begin{aligned}
\tau_E \dot{r}_{E,j} &= \frac{\Delta_E}{\pi \tau_E} + 2r_{E,j} v_{E,j} \\
\tau_E \dot{v}_{E,j} &= \bar{\eta}_E+ v_{E,j}^2 +  (\tau_E \pi r_E)^2 + I_{ext}^E + \tau_E (J_{EE}s_{E,j} - J_{EI}s_{I,j} + I_{j})\\
\tau_{s_E} \dot{s}_{E,j} &= -s_{E,j} + r_{E,j}
\end{aligned}
\right. \nonumber \\[1ex]
\text{($I_j$)}&\quad\left\{
\begin{aligned}
\tau_I \dot{r}_{I,j} &= \frac{\Delta_I}{\pi \tau_I} + 2r_{I,j} v_{I,j} \\
\tau_I \dot{v}_{I,j} &= \bar{\eta}_I + v_{I,j}^2 - (\tau_I \pi r_{I,j})^2 + I_{ext}^I + \tau_I( J_{IE}s_{E,j} -  J_{II}s_{I,j} +I_j)\\
\tau_{s_I} \dot{s}_{I,j} &= -s_{I,j} + r_{I,j}
\end{aligned}
\right. \label{eq:NetworkModel}
\end{align}
where $r_{\ell,j}$ and $v_{\ell,j}$, with $\ell \in \{E,I\}$ are the mean firing rate and mean membrane potential of population $\ell$ in node $j$, respectively. The variable $s_{\ell}$ represents the synaptic current from population $\ell$. Parameters $\tau_{\ell}$ and $\tau_{s_{\ell}}$ are the membrane and synaptic time constants, respectively. For $p,q \in \{E,I\}$, $J_{pq}$ is the connection strength from population $q$ to population $p$. 
At the single neuron description, each unit receives a different input current, which can be either static and drawn from a Cauchy distribution \cite{montbrio2015macroscopic} or Cauchy white noise \cite{clusella_exact_2024}. 
In either case, parameters $\eta_{\ell}$ and $\Delta_{\ell}$ are, respectively, the center and width of the distribution of such inputs. 

Coupling between nodes is mediated by synaptic projections originating from the excitatory populations, while inhibition acts only locally, as in similar models \cite{clusella2023complex}. In our case, these long-range excitatory inputs target both the excitatory and inhibitory populations of the receiving node. The strength of the synaptic couplings is governed by the normalized structural connectivity matrix $\tilde{\bm{W}}$. Additionally, the overall influence of these long-range interactions is modulated by a global coupling parameter $\varepsilon$, which scales the entire connectivity matrix.
Thus, the excitatory inputs from other nodes to population $\ell$ of node $j$ are given by
\[I_j=\varepsilon \displaystyle \sum_{k=1}^N\tilde{w}_{jk}s_{E,k},\]
where $\tilde{w}_{jk}$ are the synaptic weights obtained from the entries of the row-normalized structural connectivity matrix $\tilde{\bm{W}}$. 

We study the system under the influence of the external input to the excitatory population $I_{ext}^E$ and the coupling strength $\varepsilon$, leaving all the other parameters fixed. The parameters are chosen according to \cite{reyner2022phase} and are provided in Table~\ref{tab:ParamsNextGen}.
\begin{table}[h!]
    \centering
    \begin{tabular}{|c|c|c|}
    \hline 
        \textbf{Parameter} & \textbf{Meaning} & \textbf{Value}  \\
        \hline \hline
        $\tau_E,\tau_I,\tau_{s_E}, \tau_{s_I}$ & Time constants & $8,8,1,5$ ms \\
        \hline
        $\bar{\eta}_E,\bar{\eta}_I$ & Baseline constant current for E,I neurons & $-5,-5$ \\
        \hline
        $\Delta_E, \Delta_I$ & Population heterogeneity or noise strength & 1,1 \\
        \hline
        $J_{EE},J_{EI},J_{IE},J_{II}$ & Synaptic strength of connection between populations & $5,13,13,5$ \\
        \hline
        $I_{ext}^E,I_{ext}^I$ & External current impinging in E/I population & Not fixed, $0$ \\
        \hline
        $\varepsilon$ & Coupling strength between nodes & Not fixed \\
        \hline
    \end{tabular}
    \caption{Parameters of the coupled next generation neural mass model in equation \eqref{eq:NetworkModel}.}
    \label{tab:ParamsNextGen}
\end{table}

The row-normalized connectivity matrix ensures that all units in the network receive the same total input, although distributed differently across nodes. This structure guarantees the existence of an invariant manifold, with identical dynamics for all nodes. We refer to this manifold as the \emph{homogeneous invariant manifold}. We use this manifold as a reference structure to explore how instabilities of homogeneous states can emerge and generate complex spatiotemporal dynamics. To understand such collective behavior, we first study this simplified state of identical node dynamics.

\section{Homogeneous invariant manifold}\label{sec:hom}

In this section, we derive the equations governing the dynamics on the homogeneous invariant manifold and analyze their solutions and  local stability to determine which homogeneous states are stable under uniform perturbations. 

Let $\bm{y}_j=(r_{E,j},v_{E,j},s_{E,j},r_{I,j},v_{I,j},s_{I,j})$ denote the variables describing the dynamics of node $j$. Imposing that $\bm{y}_j=\bm{y}:= (r_{E},v_{E},s_{E},r_{I},v_{I},s_{I})$ for all $j=1,\ldots, N$, and using equation~\eqref{eq:HomogeneousCondition}, the incoming input for each brain region in \eqref{eq:NetworkModel} becomes\begin{equation}\label{eq:se}
    I_j=\varepsilon \displaystyle \sum_{k=1}^N\tilde{w}_{j,k}s_{E,k}  =\varepsilon s_E \displaystyle \sum_{k=1}^N\tilde{w}_{j,k} = \varepsilon s_E,
    \end{equation}
so it does not depend on the node index $j$ anymore. Replacing expression \eqref{eq:se} in the input current $I_j$ of both excitatory and inhibitory populations in~\eqref{eq:NetworkModel}, the equations describing the dynamics on the homogeneous manifold become
\begin{align}
\text{(E)}&\quad\left\{
\begin{aligned}
\tau_E \dot{r}_E &= \frac{\Delta_E}{\pi \tau_E} + 2r_E v_E, \\
\tau_E \dot{v}_E &= \bar{\eta}_E + v_E^2  - (\tau_E \pi r_E)^2
+I_{ext}^E + \tau_E [(J_{EE}+\varepsilon)s_E - J_{EI}s_I], \\
\tau_{s_E} \dot{s}_E &= -s_E + r_E,
\end{aligned}
\right. \nonumber \\[1ex]
\text{(I)}&\quad\left\{
\begin{aligned}
\tau_I \dot{r}_I &= \frac{\Delta_I}{\pi \tau_I} + 2r_I v_I, \\
\tau_I \dot{v}_I &= \bar{\eta}_I + v_I^2  - (\tau_I \pi r_I)^2 + I_{ext}^I + \tau_I [(J_{IE}+\varepsilon)s_E -  J_{II}s_I], \\  
\tau_{s_I} \dot{s}_I &= -s_I + r_I.
\end{aligned}
\right. \label{eq:HomogeneousSystem}
\end{align}

This low-dimensional system captures the dynamics of spatially homogeneous states of the full system~\eqref{eq:NetworkModel}. Local stability analysis reveals how solutions respond to uniform perturbations, that is, those that affect all nodes in an identical manner, a task we address in the next section.

\subsection{Dynamics on the homogeneous invariant manifold}

In this section, we analyze the dynamics of system~\eqref{eq:HomogeneousSystem} as external excitatory input $I_{ext}^E$ and coupling strength $\varepsilon$ vary. We use the bifurcation analysis software AUTO-07p~\cite{doedel2007auto} to identify invariant objects of the homogeneous system \eqref{eq:HomogeneousSystem} and explore their stability within the homogeneous manifold.

\begin{figure}[h!]
    \centering
    \includegraphics[width=1\linewidth]{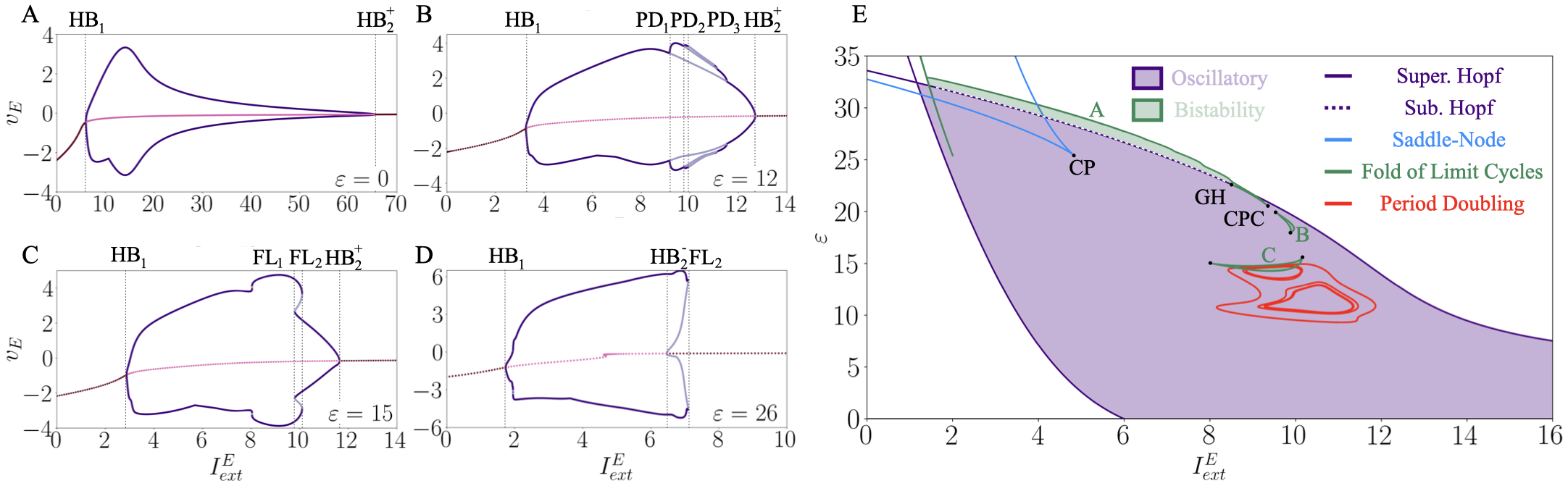}
    \caption{\footnotesize{\textbf{Bifurcation diagram of the homogeneous system~\eqref{eq:HomogeneousSystem}.} (\textbf{A-D}) One-parameter bifurcation diagram of equation~\eqref{eq:HomogeneousSystem} for varying $I_{ext}^E$ and with fixed (\textbf{A}) $\varepsilon =0$, (\textbf{B}) $\varepsilon=12$, (\textbf{C}) $\varepsilon=15$ and (\textbf{D}) $\varepsilon=26$. Colored dotted curves indicate stable equilibrium points (dark pink), unstable equilibrium point (light pink), extrema of stable limit cycle (dark purple) and extrema of unstable limit cycle (light purple). Dashed vertical lines indicate relevant bifurcation points (see text). (\textbf{E}) Two-parameter bifurcation diagram of system~\eqref{eq:HomogeneousSystem} for varying $I_{ext}^E$ and $\varepsilon$. Colored curves indicate the different bifurcations: supercritical Hopf (solid purple), subcritical Hopf (dashed purple), saddle-node (blue), fold of limit cycles (green), period doubling (red). The light purple region indicates a single periodic state, while the light green region indicates bistability (see text).
    }}
    \label{fig:BifDiagram}
\end{figure}

Figure~\ref{fig:BifDiagram}A shows the bifurcation diagram of the uncoupled model ($\varepsilon=0$ in \eqref{eq:HomogeneousSystem}) as $I_{ext}^E$ is varied. For low values of $I_{ext}^E$, there exists a unique stable state of the system given by an equilibrium point, corresponding to asynchronous activity. It loses stability via a supercritical Hopf bifurcation (HB$_1$) at $I_{ext}^E \approx 5$. Therefore, a stable limit cycle emerges and persists until $I_{ext}^E \approx 65$, where it vanishes at another supercritical Hopf bifurcation (HB$_2^+$) of the equilibrium point, which transitions back to stable. This oscillatory activity corresponds to gamma activity, with a frequency ranging 27-170 Hz.

Including coupling $\varepsilon>0$ in system~\eqref{eq:HomogeneousSystem} modifies this bifurcation scenario. In general, the oscillatory region becomes narrower and new invariant objects arise, which leads to richer and more complex behavior. Namely, for $\varepsilon=12$ (see Figure~\ref{fig:BifDiagram}B), the limit cycle loses stability through a period doubling bifurcation (PD$_1$) at $I_{ext}^E\approx 9$. This bifurcation gives rise to a new limit cycle with twice the period of the first periodic orbit. As $I_{ext}^E$ increases further, successive period-doubling bifurcations (PD$_2$, PD$_3$, ...) occur, forming a period-doubling cascade that may signal the onset of deterministic chaos \cite{strogatz2024nonlinear}. In Section~\ref{sec:PeriodDoublingChaos} we provide an in-depth analysis of this situation. 

As $\varepsilon$ is increased further, there appear regions of bistability, characterized by the simultaneous presence of two stable attractors. Namely, for $\varepsilon=15$ (see  Figure~\ref{fig:BifDiagram}C), we observe the coexistence of two stable limit cycles and one unstable limit cycle, which coalesce at a pair of fold of limit cycle bifurcations (FL$_1$ and FL$_2$). For  $\varepsilon=26$ , we encounter a region of bistability between an equilibrium point and a limit cycle. In this case, the bistability emerges through the second Hopf bifurcation (HB$_2^-$), which is now subcritical, where the equilibrium point recovers stability and an unstable limit cycle appears and persists until it coalesces with the stable limit cycle at a fold of limit cycles (FL$_2$). Also for $\varepsilon=26$, two saddle-node bifurcations of the already unstable fixed points occur within the oscillatory region (not marked in the diagram).

These dynamics can be more clearly organized in the two-parameter bifurcation diagram shown in Figure~\ref{fig:BifDiagram}E, which allows us to identify the regions of oscillatory states in the parameter space $(I_{ext}^E, \varepsilon)$.  
Fixed points, representing asynchronous activity, destabilize through Hopf bifurcations (purple curves), giving rise to gamma oscillations (purple region) that may be stable or unstable depending on parameter values. Solid (resp. dashed) purple curves correspond to supercritical (resp. subcritical) Hopf bifurcations, with the transition occurring at a generalized Hopf (GH) or Bautin codimension-$2$ bifurcation at $I_{ext}^E \approx 8.48, \varepsilon \approx 22.65$. 
Moreover, oscillations also vanish or emerge through fold of limit cycles (FLC) bifurcations (green curves), some of them colliding at a cusp of limit cycles (CPC) codimension-$2$ bifurcations.

Thus, oscillations occur in the purple and green regions enclosed by the Hopf and FLC bifurcations. In the green region A, the system presents bistability between asynchronous and oscillatory activity. There is also a tiny region (intersection of purple and green region A) showing bistability between two different types of oscillations. Indeed, in this region the system presents three limit cycles, two of them stable and one unstable, in addition to the unstable equilibrium point. Within the oscillatory purple region, there appear two islands of bistability (green regions labeled with $B$ and $C$) bounded by FLC bifurcations that coalesce at different cusps of limit cycles (indicated with black dots). For $\varepsilon$ in the range $10-15$, we identified a closed curve of period-doubling bifurcations (red), with further period-doubling bifurcations inside (red curves), suggesting a period-doubling cascade.

For large coupling values $\varepsilon$, two branches of saddle-node bifurcations (blue) appear, enclosing a region of bistability between two fixed points. These curves meet at a cusp point (CP) around $I_{ext}^E \approx 4.82, \varepsilon \approx 25.43$. At higher $\varepsilon$, additional bifurcations arise, leading to new codimension-2 bifurcations. Nonetheless, within the range explored using AUTO-07p, no stable periodic solutions were found beyond the purple and green regions depicted in Figure~\ref{fig:BifDiagram}E.

\subsection{Period doubling cascade and deterministic chaos}\label{sec:PeriodDoublingChaos} 

The bifurcation diagram shown in Figure \ref{fig:BifDiagram}E reveals a region of nested closed curves of period-doubling bifurcations, suggesting the presence of deterministic chaos. In this section, we investigate this phenomenon in detail.

\begin{figure}[h!]
    \centering
    \includegraphics[width=1\linewidth]{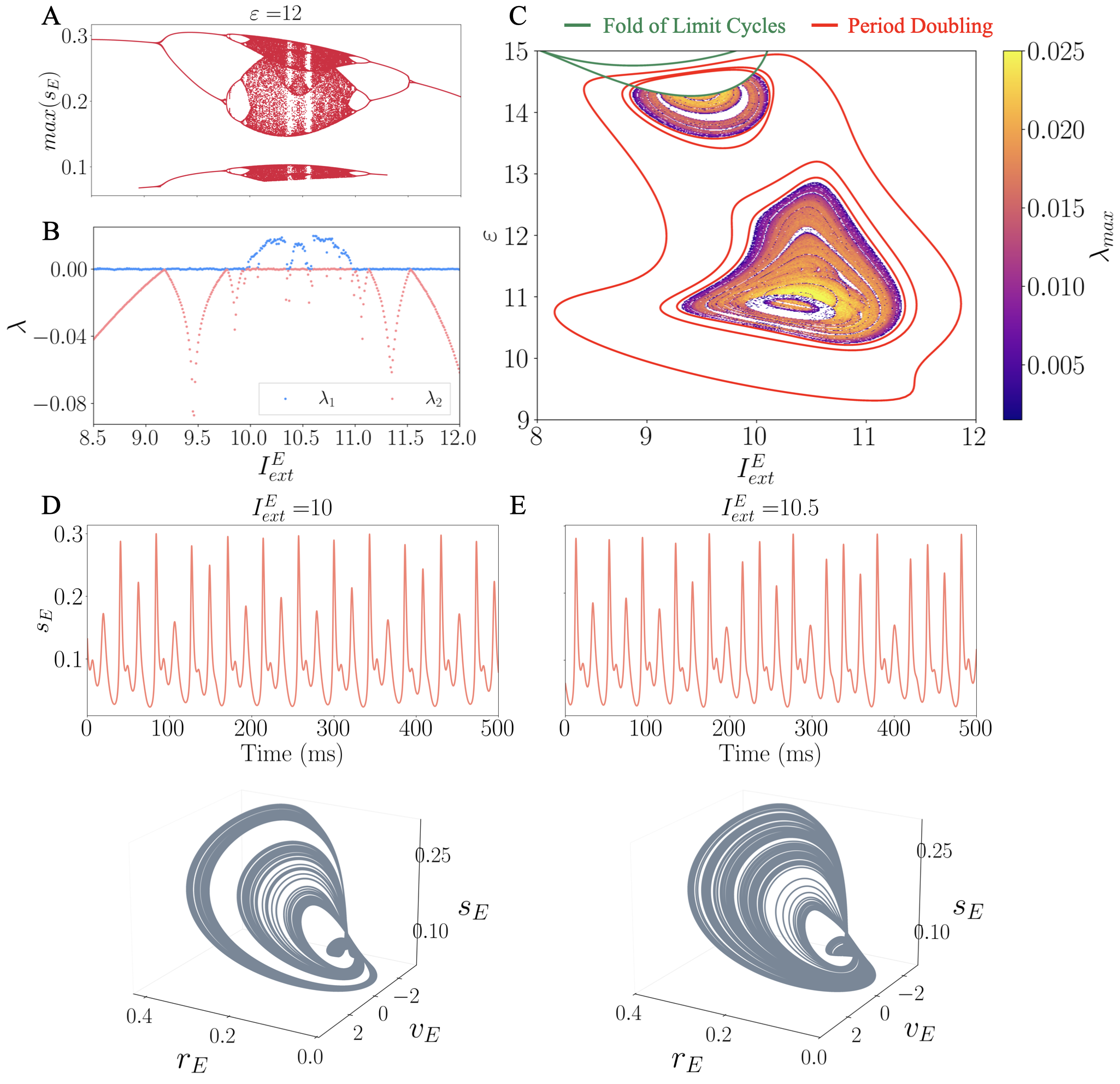}
    \caption{\footnotesize{\textbf{Analysis of the chaotic region for the homogeneous system~\eqref{eq:HomogeneousSystem}. } (\textbf{A}) Local maxima of the time course of $s_E$ variable along a trajectory of system~\eqref{eq:HomogeneousSystem} after discarding an initial transient time for fixed $\varepsilon = 12$ and varying $I_{ext}^E \in [8.5,12]$ with a discretization step $\Delta_{I_{ext}^E} = 0.001$. (\textbf{B}) Two largest Lyapunov exponents of system~\eqref{eq:HomogeneousSystem} for fixed $\varepsilon = 12$ and varying $I_{ext}^E \in [8.5,12]$ with a discretization step $\Delta_{I_{ext}^E} = 0.01$}. (\textbf{C})  Largest Lyapunov exponent of the homogeneous system varying both $\varepsilon \in [8,12]$ and $I_{ext}^E \in [9,15]$ with a discretization step $\Delta = 0.01$. The green and red curves represent, respectively, the fold of limit cycles and the period doubling bifurcation curves computed with AUTO-07p. (\textbf{D}-\textbf{E}) Simulated trajectories of the homogeneous system~\eqref{eq:HomogeneousSystem} for fixed $\varepsilon = 12$ and $I_{ext}^E=10$ (\textbf{D}) and $I_{ext}^E=10.5$ (\textbf{E}) showing convergence to the attracting solution after discarding an initial transient. The top plots show the time series of the variable  $s_E$ for $500\,\mathrm{ms}$. Bottom plots display 
    the trajectory projected onto the 3-dimensional space $(r_E, v_E, s_E)$ for $5000\,\mathrm{ms}$.} 
    \label{fig:ChaoticRegion}
\end{figure}

We begin by exploring the chaotic region at $\varepsilon=12$, which intersects the region of period-doubling curves in the two-parameter bifurcation diagram (see Figure~\ref{fig:BifDiagram}B), while varying $I_{ext}^E \in [8.5,12]$. For each value of $I_{ext}^E$, we integrate the system for $500$ ms after disregarding an initial transient. Figure~\ref{fig:ChaoticRegion}A shows the sequence of local maxima of the excitatory synapse variable $s_E$, plotted against $I_{ext}^E$.

In regions where the system exhibits stable periodic behavior, the diagram shows either one or two branches corresponding to either one or two local maxima per oscillation cycle, respectively. 
As we increase $I_{ext}^E$, the system undergoes successive period-doubling bifurcations, each doubling the number of maxima and generating a cascade that leads to deterministic chaos, reflected by the presence of dense, irregular patterns in the diagram~\cite{strogatz2024nonlinear}. The chaotic dynamics are confined to a bounded interval of the parameter space, beyond which the system returns to periodic behavior. Within the chaotic region, we also observe  bands of parameter values for $I_{ext}^E$ in which the system maintains a stable behavior, forming islands of stability.

To quantitatively characterize the chaotic dynamics,  Figure~\ref{fig:ChaoticRegion}B shows the two largest Lyapunov exponents $\lambda$ of the homogeneous system, computed numerically (see Appendix~\ref{sec:NumericalMethods}). For values of $I_{ext}^E$ corresponding to periodic oscillations, the largest Lyapunov exponent (LLE) is zero while the others remain negative. However, for values of $I_{ext}^E$ that lie in the chaotic region, the LLE becomes positive and at the critical values of $I_{ext}^E$, where period-doubling bifurcations occur, the second-largest Lyapunov exponent becomes zero.

To identify the extent of the chaotic region of the homogeneous system in the two-parameter space, Figure~\ref{fig:ChaoticRegion}C shows the LLE over $(I_{ext}^E,\varepsilon)$. The white area corresponds to the region of stable periodic oscillations, as it exhibits zero LLE. Instead, the colored areas reveal two regions of positive LLE, which fit within the curves of period-doubling bifurcations detected using AUTO-07p (red curves in Figure~\ref{fig:ChaoticRegion}C). These correspond to the 
regions of deterministic chaos in the two parameter space.
Additionally, the diagram reveals the shape and location of the islands of stability within the chaotic area.

Instances of these chaotic states are depicted in Figures~\ref{fig:ChaoticRegion}D and E, which show simulations of the homogeneous system for fixed $\varepsilon=12$ and $I_{ext}^E=10$ and $I_{ext}^E=10.5$, respectively. The upper row of each panel provides the time evolution of the excitatory synaptic variable $s_E$ over $500$ ms of a trajectory starting from initial conditions set to zero and discarding an initial transient. The lower row, depicts the same trajectory within the three-dimensional subspace spanned by the variables of the excitatory population $(r_E,v_E,s_E)$. In both cases, the system lies in the chaotic regime, exhibiting aperiodic dynamics in which the trajectory ends up densely filling a finite region of the phase space. Hence, the presence of a strange attractor becomes evident.

\section{Transverse instabilities}\label{sec:transverse}

The different dynamical regimes and stability regions uncovered so far concern only perturbations acting uniformly among all brain regions. However, perturbations that are transverse to the homogeneous invariant manifold may drive the dynamics of the large-scale brain model toward heterogeneous neural activity patterns including traveling or chaotic waves.
In this section we perform linear stability analysis of the
homogeneous fixed points and limit cycles of the full system \eqref{eq:NetworkModel}.

To this end, we apply a well-known technique of projecting the perturbation onto a suitable eigenbasis 
that decouples the high-dimensional stability problem into simpler, lower-dimensional components associated to different spatial modes. 
For fixed point solutions, this simply corresponds to the analysis of Turing instabilities in networked systems \cite{nakao_turing_2010},
whereas for limit cycles, the analogous procedure is usually known as the Master Stability Function (MSF) formalism \cite{pecora_master_1998}.
When applying this procedure, coupled neural mass models differ from other common setups based on diffusive coupling on the choice
of the eigenbasis, which here corresponds to that of the structural connectivity matrix instead of the widely used Laplacian \cite{clusella2023complex}.

Indeed, diffusion tensor imaging usually provides symmetric connectivities. 
The row-normalization of a symmetric matrix, $\tilde{\bm{W}}$, preserves some diagonalization properties of the original matrix; in particular, $\tilde{\bm{W}}$ is diagonalizable with real eigenvalues \cite{clusella2023complex}. 
By construction, its largest eigenvalue is $\Lambda_1 = 1$, with associated eigenvector $\Phi^{(1)} = (1,\ldots,1)$. 
Moreover, by the \emph{Gershgorin circle theorem} \cite{wilkinson1967}, all other eigenvalues $\Lambda_\alpha$, with $\alpha = 2,\dots,N$, are confined within the unit disk, i.e., $\Lambda_\alpha \in [-1,1]$. 
The eigenvectors of the normalized structural connectivity matrix, $\Phi^{(\alpha)}$, are also referred to as \textit{eigenmodes} in this context, since they define spatial patterns of activity across the network.

Ultimately, the linear stability analysis based on mode decomposition provides the largest exponential growth rate $\mu_{max}^{(\alpha)}$ of a perturbation acting along each eigenmode $\bm{\Phi}^{(\alpha)}$ as a function of its associated eigenvalue $\Lambda_{\alpha}$. That is, 
\begin{equation}\label{eq:MSF}
\mathcal{F}(\Lambda_{\alpha})=\mu_{max}^{(\alpha)}.
\end{equation}
We refer to this function as the dispersion relation. 
A positive value of $\mu_{max}^{(\alpha)}$ implies the instability along the $\alpha$-th eigenmode's direction.
If $\alpha>1$, this usually results in the emergence of (spatially) heterogeneous dynamics. 
On the contrary, if all $\mu_{max}^{(\alpha)}<0$ for all $\alpha=2,\dots,N$, then small heterogeneous perturbations of the homogeneous state will decay exponentially. 
In Appendix~\ref{ap:MSF} we provide a detailed derivation of the method.

\subsection{Transverse instabilities of homogeneous equilibrium points}\label{sec:TransverseInstabilitiesEqPoints}

In this section, we study the stability of homogeneous equilibrium points with respect to heterogeneous perturbations by computing the dispersion relation in equation \eqref{eq:MSF}. In this case, the growth rate $\mu_{max}^{(\alpha)}$ corresponds to the real part of the dominant eigenvalue of the  linearized system around the equilibrium. This quantity thus governs the evolution of perturbations projected onto the eigenmode $\bm{\Phi}^{(\alpha)}$, for $\alpha = 1, \ldots, N$. 

\begin{figure}[h!]
    \centering
    \includegraphics[width=1\linewidth]{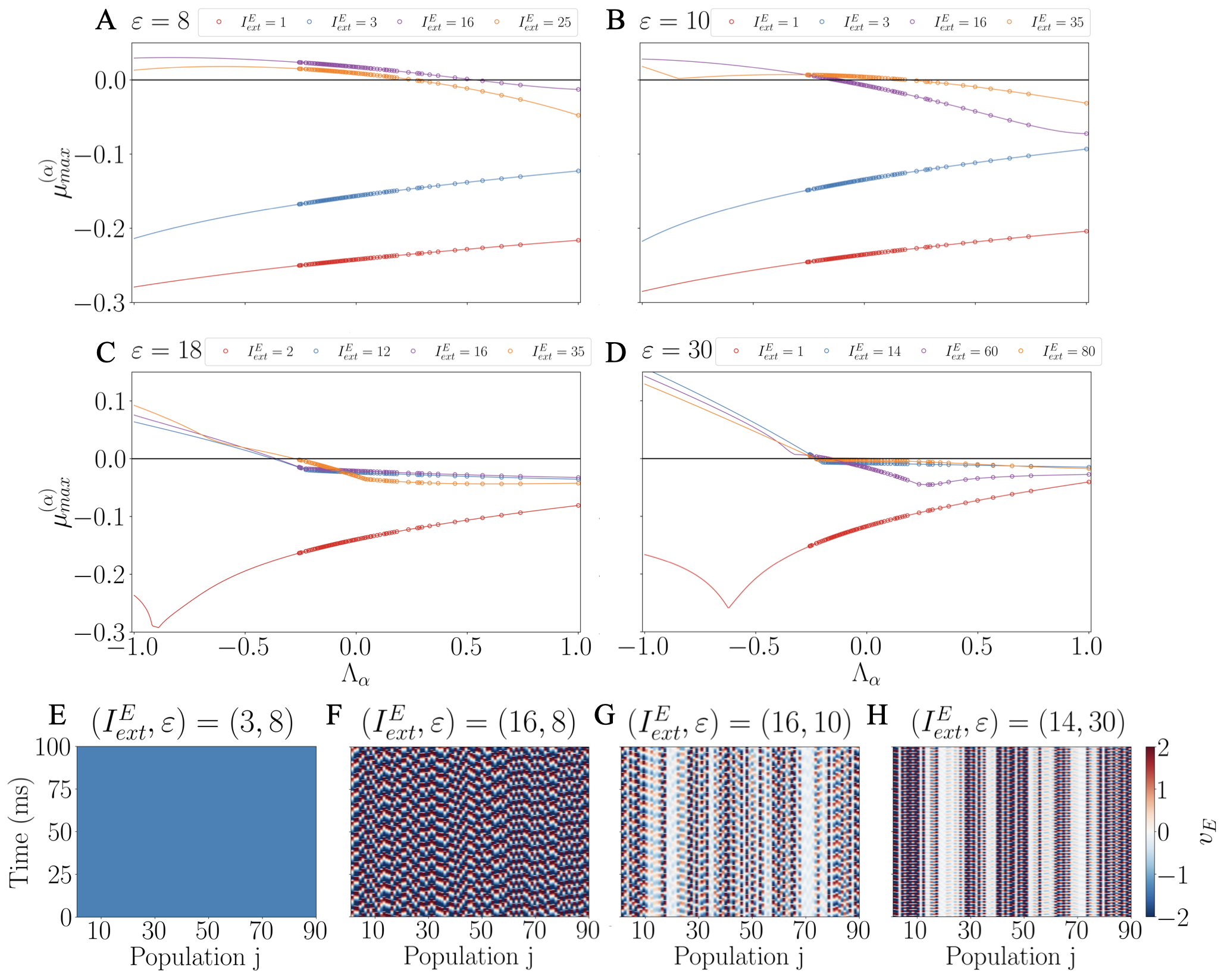}
    \caption{\footnotesize{\textbf{Transverse instabilities of homogeneous equilibrium points.} (\textbf{A}-\textbf{D}) Dispersion relation \eqref{eq:MSF} for homogeneous equilibrium points of system~\eqref{eq:NetworkModel}, showing the dependence of the growth rate $\mu_{max}^{(\alpha)}$ on the eigenvalues $\Lambda_{\alpha}$ of the structural connectivity matrix $\tilde{W}$. Parameters are fixed at $\varepsilon=8$ (\textbf{A}), $\varepsilon=10$ (\textbf{B}), $\varepsilon=18$ (\textbf{C}) and $\varepsilon=30$ (\textbf{D}) with different values of $I_{ext}^E$ indicated in each panel. Continuous curves are obtained by considering a continuous range of $\Lambda_{\alpha} \in [-1,1]$. (\textbf{E}-\textbf{H}) Numerical simulations of the large-scale brain model~\eqref{eq:NetworkModel} for different combinations of the parameters $(I_{ext}^E,\varepsilon)$, each starting from a homogeneous equilibrium with a small random perturbation. Each plot shows the evolution of the  variable $v_E$ (color-coded) for each node $j=1,\ldots,N$ ($x$-axis) over 100 ms ($y$-axis) after discarding an initial transient.}}
    \label{fig:MSF_EqPoints}
\end{figure}

Figure~\ref{fig:MSF_EqPoints}A-D shows the dispersion relation \eqref{eq:MSF} of the stable equilibria of the homogeneous system~\eqref{eq:HomogeneousSystem} for selected parameter combinations $(I_{ext}^E, \varepsilon)$.
Since the uniform eigenmode, associated to  $\Lambda_{\alpha} = 1$, corresponds to perturbations along the homogeneous manifold, the largest growth rate for perturbations along this direction is negative. The real part of the eigenvalues associated to the remaining eigenmodes may be positive or negative.

For $\varepsilon=8$ (Figure~\ref{fig:MSF_EqPoints}A) and $\varepsilon=10$ (Figure~\ref{fig:MSF_EqPoints}B), low values of $I_{ext}^E$ yield negative growth rates across all the eigenmodes, indicating stability of the homogeneous state. However, as $I_{ext}^E$ increases, the dispersion curves are shifted upwards and the growth rates associated to eigenmodes with low structural eigenvalues $\Lambda_\alpha$ become positive, leading to transverse instabilities of the homogeneous state. As $\varepsilon$ increases to $\varepsilon=18$ (Figure~\ref{fig:MSF_EqPoints}C), the dispersion relation becomes negative for all eigenmodes across the different values of $I_{ext}^E$ explored. Therefore, the homogeneous equilibrium point is transversely stable for all these parameter combinations. Finally, for $\varepsilon=30$ (Figure~\ref{fig:MSF_EqPoints}D), high values of $I_{ext}^E$ lead to some eigenmodes becoming unstable once again, indicating loss of stability for the equilibrium point.

To evaluate the predictions of the linear stability analysis, we compare them with simulations of the large-scale brain model (Figures~\ref{fig:MSF_EqPoints}E–H) for representative parameter combinations. All simulations start from a homogeneous equilibrium with a small random perturbation, and we show the last $100$ms of a $60$s run. For $\varepsilon=8$, $I_{ext}^E=3$ (blue circles in Figure~\ref{fig:MSF_EqPoints}A), the dispersion relation shows that all modes decay (negative growth rate), and the network dynamics decay to the stable homogeneous equilibrium, as seen in Figure~\ref{fig:MSF_EqPoints}E.
In contrast, Figures~\ref{fig:MSF_EqPoints}F-H correspond to parameter combinations with a  dispersion relation where certain modes have positive growth rates, leading to heterogeneous spatiotemporal dynamics. It is  remarkable that in all cases, destabilization of the homogeneous equilibrium point produces oscillatory patterns with features specific to each parameter set. For example, for $\varepsilon=10$ (Figure~\ref{fig:MSF_EqPoints}G) and $\varepsilon=30$ (Figure~\ref{fig:MSF_EqPoints}H), some nodes exhibit oscillations with significantly lower amplitude than others, behavior that is not observed for $\varepsilon=8$ (see Figure~\ref{fig:MSF_EqPoints}F). In Section~\ref{sec:carac}
we characterize some of these complex states in detail. 

\subsection{Transverse instabilities of homogeneous periodic orbits}\label{sec:TransverseInstabilitiesOscillatory}

We analyze the stability of homogeneous periodic orbits with respect to heterogeneous perturbations by computing the dispersion relation in equation~\eqref{eq:MSF}, which in the case of periodic orbits is  also known as the Master Stability Function (MSF) \cite{pecora_master_1998}. In this case, the growth rate $\mu_{max}^{(\alpha)}$ corresponds to the real part of the largest Floquet exponent of the periodic orbit, which governs the evolution of perturbations applied along the 
$\alpha$-th eigenmode. In
Appendix~\ref{ap:MSF_num} we provide details of the numerical computation. 

\begin{figure}[h!]
    \centering
    \includegraphics[width=1\linewidth]{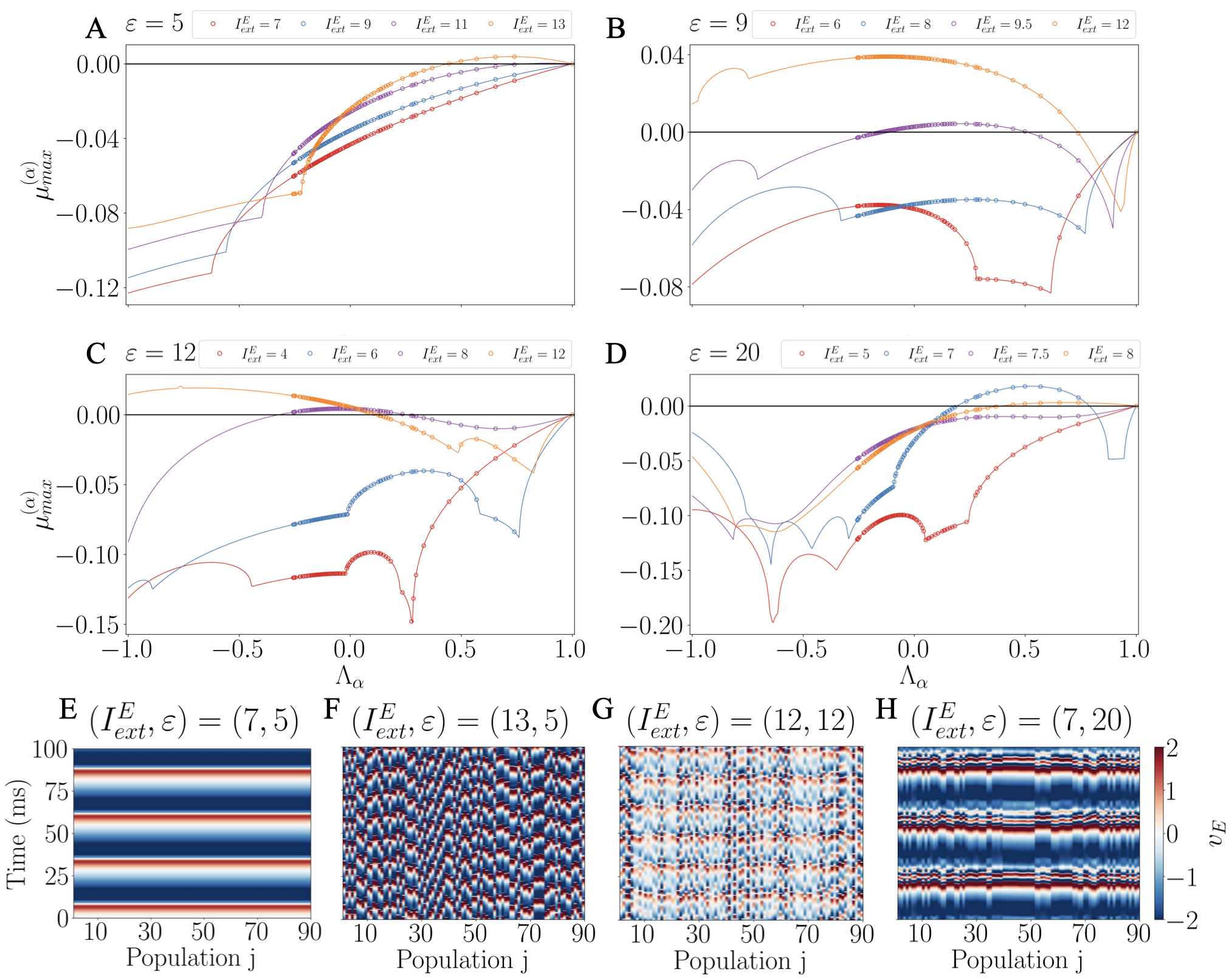}
    \caption{\footnotesize{\textbf{Transverse instabilities of homogeneous periodic solutions.} (\textbf{A}-\textbf{D}) Dispersion relation, known as Master Stability Function, for homogeneous periodic orbits of system~\eqref{eq:NetworkModel}, showing the dependence of the real part of the largest Floquet exponent $\mu_{max}^{(\alpha)}$ on the structural connectivity eigenvalues $\Lambda_{\alpha}$. Parameters are fixed at $\varepsilon=5$ (\textbf{A}), $\varepsilon=9$ (\textbf{B}), $\varepsilon=12$ (\textbf{C}) and $\varepsilon=20$ (\textbf{D}) with different values of $I_{ext}^E$ indicated in each panel. Continuous curves are obtained by considering a continuous range of $\Lambda_{\alpha} \in [-1,1]$. (\textbf{E}-\textbf{H}) Numerical simulations of the large-scale brain model~\eqref{eq:NetworkModel} for different combinations of parameters $(I_{ext}^E,\varepsilon)$, each starting from a point on a homogeneous periodic orbit with a small random perturbation. Each plot shows the evolution of the  variable $v_E$ (color-coded) for each node $j=1,\ldots,N$ ($x$-axis) over 100 ms ($y$-axis) after discarding an initial transient.}}
    \label{fig:MSF_PeriodOrbits}
\end{figure}

Figure~\ref{fig:MSF_PeriodOrbits}A-D shows the MSF for stable periodic orbits of the homogeneous system~\eqref{eq:HomogeneousSystem} for selected parameter combinations $(I_{ext}^E, \varepsilon)$. 
The largest growth rate associated to homogeneous perturbations ($\alpha=1$) is $\mu_{max}^{(1)}=0$, according to Floquet theory.
The Floquet exponents associated to the remaining eigenmodes may be positive, negative or zero.

For $\varepsilon=5$ (Figure~\ref{fig:MSF_PeriodOrbits}A), increasing $I_{ext}^E$ shifts the dispersion curves upward smoothly, preserving their shape. Notably, for larger $\varepsilon$ (Figures~\ref{fig:MSF_PeriodOrbits}B-D), the curves change shape more markedly with $I_{ext}^E$. For instance, at $\varepsilon=9$ and $I_{ext}^E = 12$ (orange circles in Figure~\ref{fig:MSF_PeriodOrbits}B), the first unstable mode corresponds to $\alpha=2$, whereas for $I_{ext}^E = 9.5$ (purple circles) the instability occurs first for $\alpha = 5$.
At $\varepsilon=20$ (Figure~\ref{fig:MSF_PeriodOrbits}D), the curves display non-monotonic behavior as $I_{ext}^E$ varies. In addition, the number of eigenmodes that become unstable varies for each parameter combination. Overall, the non-trivial shape of the dispersion relation illustrates the complexity of the bifurcation structure underlying transverse instabilities of the oscillatory states.

The high complexity exhibited by the MSF in our model for different parameter combinations $(I_{ext}^E,\varepsilon)$ may influence the types of patterns that arise when the homogeneous state destabilizes. To illustrate so, Figure~\ref{fig:MSF_PeriodOrbits}E-H shows simulations of the large-scale brain model for representative parameter combinations.

For $\varepsilon=5$ and $I_{ext}^E = 7$ (red circles in Figure~\ref{fig:MSF_PeriodOrbits}A), the dispersion relation shows a negative growth rate for all the structural eigenvalues and the network dynamics decays to the homogeneous periodic state (see Figure~\ref{fig:MSF_PeriodOrbits}E). In contrast, for $I_{ext}^E = 13$ (orange circles in Figure~\ref{fig:MSF_PeriodOrbits}A), the dispersion curve presents some modes with positive Floquet exponents, leading to heterogeneous spatiotemporal dynamics (see Figure~\ref{fig:MSF_PeriodOrbits}F).
The patterns emerging from destabilization of the homogeneous oscillatory state vary with parameters. Examples of other spatiotemporal patterns are shown in Figures~\ref{fig:MSF_PeriodOrbits}G–H. For instance, the dynamics for $\varepsilon = 5$ (Figure~\ref{fig:MSF_PeriodOrbits}F) and $\varepsilon = 12$ (Figure~\ref{fig:MSF_PeriodOrbits}G) appears rather heterogeneous, while for $\varepsilon = 20$ (Figure~\ref{fig:MSF_PeriodOrbits}H), it retains some structured organization among some nodes. Section~\ref{sec:carac} provides an in-depth characterization of these dynamics.

\subsection{Bifurcation diagram for transverse instabilities}\label{sec:BifTransverse}

In this section we thoroughly analyze the regions in the parameter space $(I_{ext}^E,\varepsilon) \in [0,16]\times[0,30]$ where we encounter transverse instabilities of the homogeneous states for both equilibrium points and periodic orbits. 

\begin{figure}[h!]
    \centering
    \includegraphics[width=0.8\linewidth]{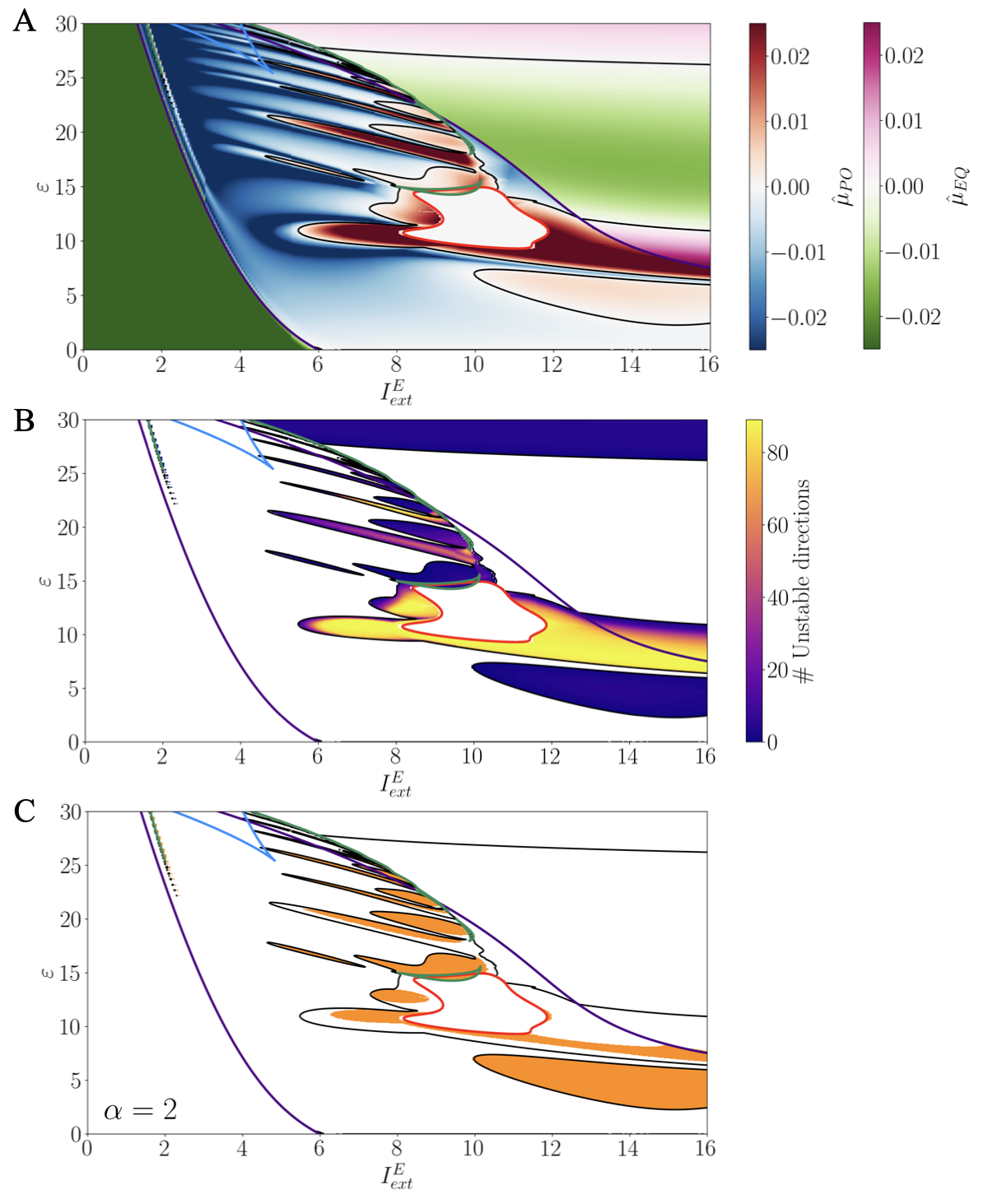}
    \caption{\footnotesize{\textbf{Bifurcation diagram of transverse instabilities.} (\textbf{A}) Largest growth rate, $\hat{\mu}$, obtained from the dispersion relation~\eqref{eq:MSF} according to formula in equation~\eqref{eq:muhat}, with values for equilibrium points ($\hat{\mu}_{EQ}$) and periodic orbits ($\hat{\mu}_{PO}$) shown using different color gradients.
    (\textbf{B}) Number of unstable directions from the dispersion relation~\eqref{eq:MSF}. (\textbf{C}) Regions (in orange) where the second eigenmode, $\Lambda_{2}$, destabilizes. In all panels, bifurcation curves of homogeneous system are overlaid using the same color code as in Figure~\ref{fig:BifDiagram}E.}} 
    \label{fig:MSFParamSpace}
\end{figure}

For each pair of parameters $(I_{ext}^E,\varepsilon)$ we compute the dispersion relation corresponding to the homogeneous solution. Figure~\ref{fig:MSFParamSpace}A shows, for each pair, the largest growth rate $\hat{\mu}$ among all eigenmodes (excluding the uniform one), that is,
\begin{equation}\label{eq:muhat}
\hat{\mu} := \max_{\alpha=2,\ldots,N}\left \{\mu_{max}^{(\alpha)}\right\}.
\end{equation}
Two distinct color gradients are used to distinguish whether the underlying solution is a fixed point, $\hat{\mu}_{EQ}$,
or a periodic orbit, $\hat{\mu}_{PO}$. For regions with bistability between homogeneous equilibrium and periodic orbits see supplementary figure \ref{fig:Bistability}. Black curves delimit the regions of transverse instabilities
and color curves correspond to bifurcations of the homogeneous system, as in Figure~\ref{fig:BifDiagram}E. 

Regarding homogeneous equilibrium points, the region to the left of the Hopf bifurcation HB$_1$ corresponds to a transversely stable equilibrium, as indicated by the strongly negative $\hat{\mu}_{EP}$.
In contrast, to the right of the Hopf bifurcation HB$_2$, the equilibrium point becomes transversely unstable in two parameter regions (warm colors), which are separated by an intermediate band where it is stable (cold colors). 

These findings differ from those reported in~\cite{clusella2023complex} for the Jansen-Rit model, where all stable equilibrium points in the homogeneous invariant manifold were found to be stable with respect to transverse perturbations.

Concerning homogeneous periodic orbits, the parameter regions where they become transversely unstable show several noteworthy features. In particular, the area with positive $\hat{\mu}_{PO}$ (colored in red in Figure~\ref{fig:MSFParamSpace}A) displays several distinct lobes characterized by different magnitudes of $\hat{\mu}_{PO}$. For instance, the large lower-right lobe displays smaller values of $\hat{\mu}_{PO}$ compared to the large central lobe, which surrounds the region where the homogeneous system exhibits chaos. Toward the upper left of the oscillatory region, additional lobes appear, alternating in the magnitude of $\hat{\mu}_{PO}$. The more elongated lobes tend to correspond to higher values of $\hat{\mu}_{PO}$, whereas shorter and rounder lobes generally present lower values. Interestingly, in the upper part of the diagram, regions with unstable periodic orbit are consistently surrounded by regions with strongly negative $\hat{\mu}_{PO}$.
 
This intrincate structure of transverse bifurcations is significantly different of the one reported in \cite{clusella2023complex} for the Jansen-Rit model, where only a single, well-defined region with unstable directions was observed. 

Figure~\ref{fig:MSFParamSpace}B shows the number of unstable directions (positive values in the dispersion relation) within the regions with transverse instabilities. For equilibrium points, we observe a high number of unstable directions (up to 89) within the lower unstable region, in contrast to the few (less than 10) unstable modes of the upper region. For oscillations, the lower lobe exhibits few unstable directions, less than 10. In contrast, the central lobe surrounding the chaotic region of the homogeneous state, shows the highest number of unstable directions, reaching up to 89 in the central and lower-right area. As for the upper lobes, the elongated ones display a higher number of unstable directions, around 30–60, while the shorter ones seem to present less unstable directions.

Finally, in Figure~\ref{fig:MSFParamSpace}C, we investigate the mechanisms underlying the shape of the regions with transverse instabilities. We highlight (in orange) the regions where the second eigenmode ($\alpha = 2$) becomes unstable. This reveals that, in the region with equilibrium points, the instabilities are driven by modes associated with lower values of $\Lambda_\alpha$. In contrast, within the oscillatory region, the second mode is the first to destabilize over a large portion of the unstable region. For instance, it completely fills the lower lobe, indicating that here the bifurcation curve is determined by the destabilization of the second eigenmode. However, some parts of the bifurcation curve remain uncovered, thus destabilization occurs through lower values of $\Lambda_\alpha$ (see Supplementary Figure \ref{fig:Eigenmodes}). Again, this behavior highly contrasts with the findings of \cite{clusella2023complex}, where transverse instabilities always arose from the destabilization of the second eigenmode ($\alpha=2$).

\section{Heterogeneous spatiotemporal
patterns in the large-scale brain model}\label{sec:carac}

\label{sec:ExplorationHeterogeneousPatterns}

The analysis in the previous sections identified parameter regions where synchrony breaks down, but it did not characterize the dynamics that follow. 
In this section, we explore the complex spatiotemporal patterns that emerge beyond the loss of stability of homogeneous solutions. To this end, we adopt two complementary approaches. First, we compute the Lyapunov exponents of system~\eqref{eq:NetworkModel} to distinguish chaotic dynamics, particularly high-dimensional chaos, from regular solutions. Then, we perform a Fourier spectral analysis to examine the frequency content of the emergent patterns. This allows us to better characterize whether the system exhibits coherent oscillations or broadband chaotic activity.
Indeed, the emergence of high-dimensional chaos aligns with recent studies that highlight the role of turbulent, complex dynamics as an indicator of healthy brain function~\cite{deco_turbulent-like_2020}.

\subsection{Lyapunov exponents analysis}\label{sec:CharacterizationSpatiotemporalPatterns}

When the homogeneous state destabilizes, the system can exhibit a rich variety of spatiotemporal patterns. Lyapunov exponents (LE) 
quantify the growth rates of infinitesimal perturbations of a system's attractor, thus providing a rigorous tool to analyze the nature of these states. 
We compute the largest 90 Lyapunov exponents $\lambda_1\leq \ldots \leq\lambda_{90}$ of system \eqref{eq:NetworkModel} using the Julia package ChaosTools (see Appendix \ref{sec:NumericalMethods} for details).
By examining the Largest Lyapunov exponent (LLE), $\lambda_1$, we can characterize the dynamics
 as chaotic ($\lambda_1>0$), periodic or quasiperiodic ($\lambda_1 = 0$), or stationary ($\lambda_1<0$).
 
 In the case of chaotic dynamics, additional exponents provide valuable information regarding the state's complexity.
 In particular, we use them to compute the Kaplan-Yorke dimension~\cite{Kaplan1979,pikovsky2016lyapunov}, which provides an estimation of the fractal dimension of a strange attractor by
 measuring how much \textit{volume} in state space is actively being stretched and filled out by the dynamics. This measure is computed as,
\begin{equation}\label{eq:KY}
 D_{KY} = j+\displaystyle \frac{\sum_{i=1}^j\lambda_i}{|\lambda_{j+1}|},
\end{equation}
where $j$ is the largest integer such that $$\displaystyle \sum_{i=1}^j\lambda_i \geq 0 \qquad \text{ and } \qquad \displaystyle \sum_{i=1}^{j+1}\lambda_i < 0.$$
Note that if $\lambda_i < 0$ for all $i$, then $D_{KY} = 0$, indicating contraction in all directions and convergence of the dynamics to a fixed point. 

\begin{figure}[h!]
    \centering
    \includegraphics[width=0.8\linewidth]{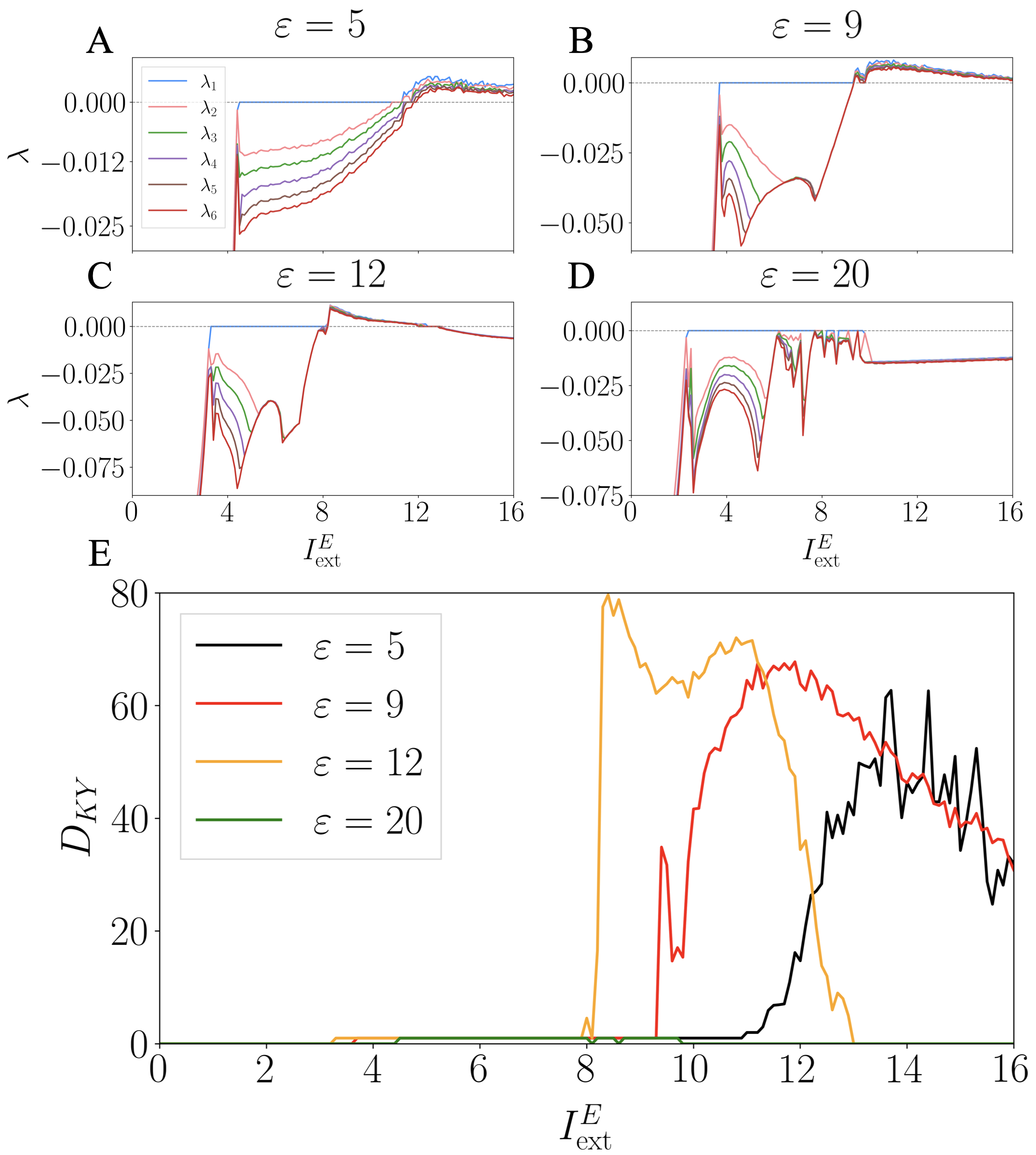}
    \caption{\footnotesize{\textbf{Lyapunov exponents of the large-scale brain model.} (\textbf{A}-\textbf{D}) First six largest Lyapunov exponents of system~\eqref{eq:NetworkModel} as a function of $I_{ext}^E \in [0,16]$ with discretization step $\Delta_{I_{ext}^E}=0.1$ and fixed $\varepsilon=5$ (\textbf{A}), $\varepsilon=9$ (\textbf{B}), $\varepsilon=12$ (\textbf{C}) and $\varepsilon=20$ (\textbf{D}). For each parameter combination, the $90$ largest Lyapunov exponents were computed starting from initial conditions near the homogeneous state with a small random perturbation. (\textbf{E}) Kaplan-Yorke dimension of the chaotic attractor, computed according to equation~\eqref{eq:KY}. 
    }}
    \label{fig:LyapunovExponentsLSModel}
\end{figure}

Figure~\ref{fig:LyapunovExponentsLSModel}A-C displays the first six Lyapunov exponents (out of the 90 computed) of system~\eqref{eq:NetworkModel} for representative values of the coupling strength $\varepsilon$ and varying $I_{ext}^E$.
We observe the onset of oscillatory activity, when the largest Lyapunov exponent passes from being negative to being zero, and the onset of chaotic dynamics, when it becomes positive. Notably, for $\varepsilon=5$, $\varepsilon=9$, and $\varepsilon=12$ (Figures~\ref{fig:LyapunovExponentsLSModel}A-B-C, respectively), the onset of chaotic dynamics coincides closely with the emergence of transverse instabilities of periodic orbits studied in Sections~\ref{sec:TransverseInstabilitiesOscillatory} and~\ref{sec:BifTransverse}. This can be observed by comparing with a horizontal cut at the corresponding $\varepsilon$ in Figure~\ref{fig:MSFParamSpace}. Remarkably, for $\varepsilon = 9$ (Figure~\ref{fig:LyapunovExponentsLSModel}B), the largest Lyapunov exponent remains positive at high values of $I_{ext}^E$, where we have detected transverse instabilities of the equilibrium points (see Figure~\ref{fig:MSFParamSpace}A). For $\varepsilon=20$ (Figure~\ref{fig:LyapunovExponentsLSModel}D), the LLE remains zero over the majority of the transverse instability region, indicating that the destabilization of the homogeneous state gives rise to heterogeneous periodic dynamics, in this case, a traveling wave.

To better identify the features of the chaotic attractor, Figure~\ref{fig:LyapunovExponentsLSModel}E displays the Kaplan-Yorke dimension of the system's attractor. High dimensional chaotic dynamics arise at $\varepsilon=5$, $9$, and $12$, with attractors with dimension almost 80 in some cases.
For $\varepsilon=9$ and $12$, this complexity can be traced back to the high number of unstable directions
of the homogeneous state (see Figure~\ref{fig:MSFParamSpace}B). 
Nonetheless, for $\varepsilon=5$ the emergence of high-dimensional dynamics may be less expected, as the underlying
homogeneous state contains less than 10 unstable directions (see Figure \ref{fig:MSFParamSpace}B). 
These results show the ubiquity of high-dimensional chaos emerging from transverse instabilities in the parameter space, highlighting thus the dynamical complexity of the brain model. 

\subsection{Frequency analysis}
To further characterize the different heterogeneous oscillatory patterns arising from transverse instabilities of both equilibrium points and periodic orbits, we analyze the frequency components present in the simulated signals. We selected some representative parameter combinations for which the homogeneous solutions are unstable for different eigenvalue characteristics. 

\begin{figure}[h!]
    \centering
    \includegraphics[width=0.7\linewidth]{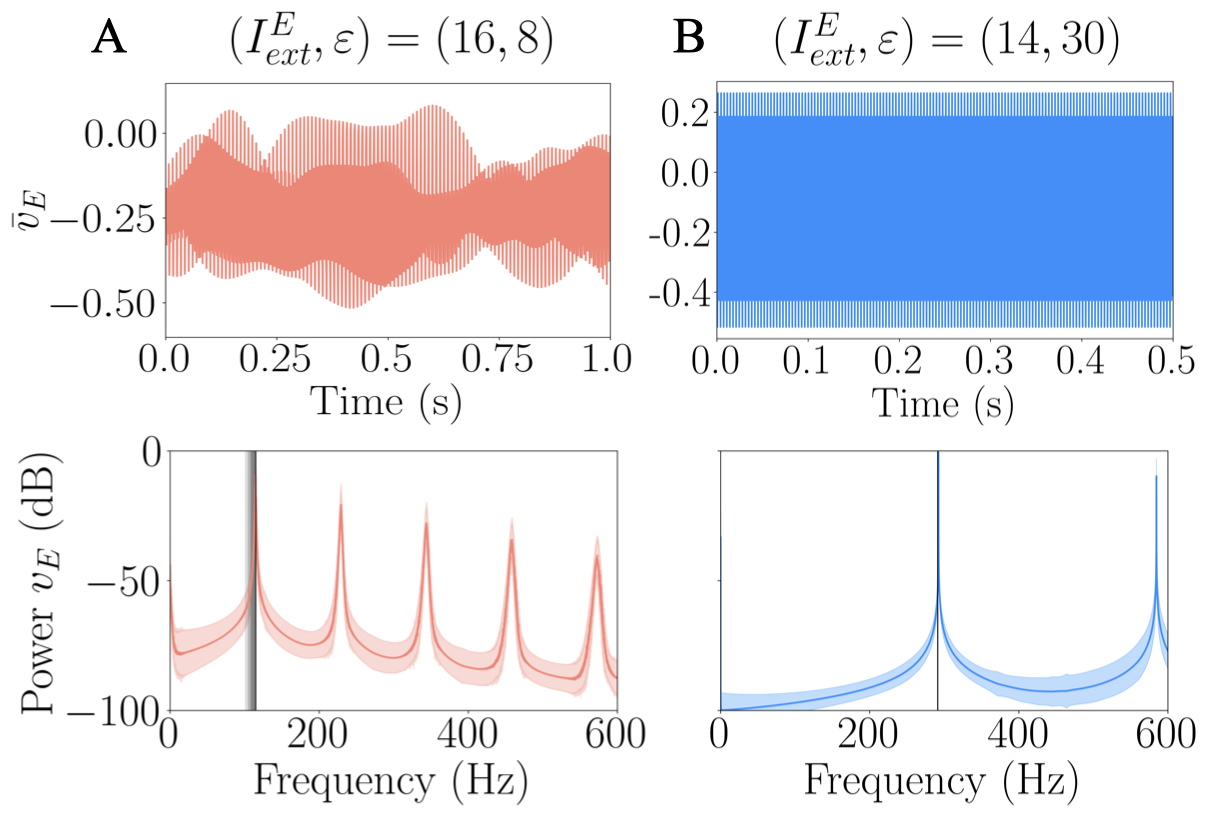}
    \caption{\footnotesize{\textbf{Frequency spectrum of  patterns emerging from homogeneous unstable equilibria.} For the parameter combinations (\textbf{A}) $(I_{ext}^E,\varepsilon)=(16,8)$ and (\textbf{B}) $(I_{ext}^E,\varepsilon)=(14,30)$, top panels display time series of the mean excitatory membrane potential $\bar{v}_{E,j}(t)$ computed across nodes for solutions of the large-scale brain model~\eqref{eq:NetworkModel} starting near the unstable homogeneous equilibrium. Each simulation was run for $60$ seconds with a time step of $0.01$, discarding the first $10$ seconds to remove transients. Bottom panels display the average frequency spectrum across all nodes (solid curve) with the corresponding standard deviation (shaded area). Vertical gray lines correspond to the frequencies obtained from the imaginary part of the eigenvalues $\mu_{max}^{(\alpha)}$ with positive real part.
    The sampling frequency is $f_s = 1\times 10^5$ samples per second and, yielding a frequency resolution of $\Delta f = 0.02$ Hz.
    }}
    \label{fig:Spectrum_EqPoints}
\end{figure}

\begin{figure}
    \centering
    \includegraphics[width=1\linewidth]{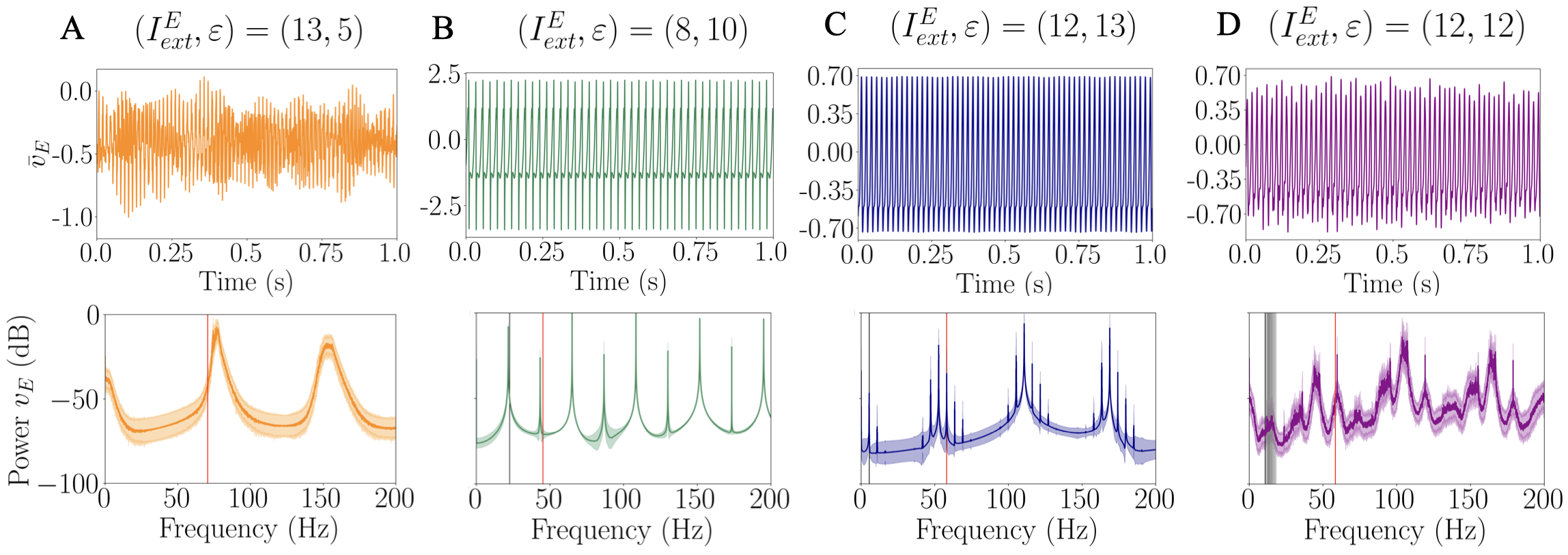}
    \caption{\footnotesize{\textbf{Frequency spectrum of patterns emerging from homogeneous unstable periodic orbits.} Analogous to Figure \ref{fig:Spectrum_EqPoints}, for the parameter combinations (\textbf{A}) $(I_{ext}^E,\varepsilon)=(13,5)$, (\textbf{B}) $(I_{ext}^E,\varepsilon)=(8,10)$, (\textbf{C}) $(I_{ext}^E,\varepsilon)=(12,13)$ and (\textbf{D}) $(I_{ext}^E,\varepsilon)=(12,12)$, (top) time series of the mean excitatory membrane potential $\bar{v}_{E}(t)$ computed across nodes for solutions of the large-scale brain model~\eqref{eq:NetworkModel} starting near the unstable homogeneous periodic orbit. Each simulation was run for $60$ seconds with a time step of $0.01$, discarding the first $10$ seconds to remove transients. Bottom panels display the average frequency spectrum across all nodes (solid curve) with the corresponding standard deviation (shaded area). The vertical red line corresponds to frequency of the homogeneous unstable periodic orbit, while the vertical gray lines correspond to the frequencies obtained from the imaginary part of the complex Floquet exponents $\mu_{max}^{(\alpha)}$ with positive real part.
    The sampling frequency is $f_s = 10^5$ samples per second and, yielding a frequency resolution of $\Delta f = 0.02$ Hz.}}
    \label{fig:Spectrum_PeriodOrbits}
\end{figure}

The signals we analyze are obtained integrating the large-scale brain model \eqref{eq:NetworkModel} for 60 seconds from initial conditions near the unstable homogeneous state (we added a small random heterogeneous perturbation to the homogeneous solution). For each simulation, we compute the frequency spectrum of each node's signal, namely the excitatory membrane potential $v_{E}$. Figures \ref{fig:Spectrum_EqPoints} and \ref{fig:Spectrum_PeriodOrbits} show the average of the spectra across all nodes, and the corresponding standard deviation. We also depict time series of the mean excitatory membrane potential, that is,
\begin{equation}
    \bar{v}_E(t) = \frac{1}{N}\displaystyle \sum_{j=1}^Nv_{E,j}(t). \nonumber
\end{equation}
These simulations show a wide range of behaviors, including heterogeneous periodic patterns (Figure \ref{fig:Spectrum_EqPoints}B, Figure \ref{fig:Spectrum_PeriodOrbits}B), amplitude modulated periodic patterns (Fig. \ref{fig:Spectrum_PeriodOrbits}C), and chaotic dynamics (Fig \ref{fig:Spectrum_EqPoints}A and \ref{fig:Spectrum_PeriodOrbits}A,D). We now discuss all these cases in detail and examine whether the peaks observed at prominent frequencies in the spectrum can be associated with the unstable directions of the homogeneous solution.

Recall that when an equilibrium point destabilizes via a supercritical Hopf bifurcation (a pair of complex eigenvalues crosses the imaginary axis) a limit cycle is created whose frequency close to the bifurcation can be approximated by the imaginary part of the corresponding eigenvalues according to the formula
\begin{equation}
f_{\mu}=\dfrac{\text{Im}(\mu)}{2 \pi} \, \text{(kHz)}.
\end{equation}
Similarly, when a periodic orbit loses stability due to a pair of complex-conjugate Floquet exponents crossing the imaginary axis, a torus (Neimark–Sacker) bifurcation typically occurs under generic conditions \cite{Kuznetsov2023}. At this bifurcation point, a smooth two-dimensional invariant torus emerges from the original periodic orbit. The resulting dynamics on it (close to the bifurcation) are quasiperiodic and characterized by the frequency of the unstable periodic orbit and the one associated to the imaginary part of the critical Floquet exponents, as described by the formula above. As system parameters vary further, the invariant torus may break down, leading to chaotic dynamics (via the so-called torus breakdown route to chaos \cite{Ruelle1971}).

In the case that the Floquet exponents are complex but the corresponding Floquet multipliers are negative real, a flip (period-doubling) bifurcation occurs: the stability of the original orbit is transferred to a new orbit with twice the period. As the system undergoes successive bifurcations, increasingly complex dynamics can develop, eventually leading to chaotic behavior.

For each parameter combination, we consider the unstable homogeneous solution and we examine both the number of unstable eigenvalues (those with positive real part) and whether these eigenvalues are real or complex. In cases where they are complex, we also consider their imaginary parts and compute the frequency according to the formula above. This information provides useful intuition for interpreting the frequency spectrum of the resulting spatiotemporal patterns, particularly for parameter values close to the bifurcation that yield the destabilization of the homogeneous solution.

We consider first the case of homogeneous equilibria (Figure~\ref{fig:Spectrum_EqPoints}). For parameter values $(I_{ext}^E, \varepsilon) = (16,8)$ (Figure~\ref{fig:Spectrum_EqPoints}A), the homogeneous solution exhibits a large number of unstable directions (86 eigenvalues with positive real parts, but with similar imaginary parts), yielding high-dimensional chaos (see Movie~\movieref{movie:Movie1}). The time series of the emerging solution reveals dynamics occurring on at least two different timescales: a fast oscillation modulated by a slower, irregular aperiodic envelope (Figure \ref{fig:Spectrum_EqPoints}A, top). Interestingly, the imaginary parts of the destabilizing eigenvalues capture the fast frequency component visible in the peaks of the power spectrum (gray lines in Figure \ref{fig:Spectrum_EqPoints}A, bottom), while the presence of significant power at very low frequencies reflects the slower modulation.

In contrast, for parameter values $(I_{ext}^E, \varepsilon) = (14,30)$ (Figure \ref{fig:Spectrum_EqPoints}B), the homogeneous equilibrium has a much smaller number of unstable directions (only six) but, as before they are complex with similar imaginary parts. In this case the time series exhibits a periodic orbit (Figure \ref{fig:Spectrum_EqPoints}B top; see also Movie~\movieref{movie:Movie2}) with a single dominant fast oscillatory component, as reflected in the power spectrum, whose fundamental frequency is well captured by the imaginary parts of the destabilizing eigenvalues (Figure \ref{fig:Spectrum_EqPoints}B, bottom).

Next, we explore patterns emerging from the destabilization of a homogeneous periodic orbit (Figure \ref{fig:Spectrum_PeriodOrbits}). 
For parameter values $(I_{ext}^E, \varepsilon) = (13,5)$ (Figure \ref{fig:Spectrum_PeriodOrbits}A), the number of unstable directions of the homogeneous solution is 5 and all the destabilizing Floquet exponents are real. Remarkably, despite the low number of unstable directions, this configuration leads to a chaotic solution similar to the one in Figure \ref{fig:Spectrum_EqPoints}A (see Movie~\movieref{movie:Movie3}). The power spectrum exhibits broad peaks around the fast frequency which, in this case, is well captured by the frequency of the unstable homogeneous periodic solution (red vertical line), along with significant power at low frequencies (Figure \ref{fig:Spectrum_PeriodOrbits}A bottom).

For parameter values $(I_{ext}^E, \varepsilon) =(8,10)$ the system exhibits a regular periodic oscillation (Figure \ref{fig:Spectrum_PeriodOrbits}B top; see also Movie~\movieref{movie:Movie4}). In this case, the homogeneous solution has 85 unstable directions, with complex Floquet exponents corresponding to negative real Floquet multipliers. These indicate that the associated eigenvectors are flipped by the Poincaré map, a signature of a period-doubling bifurcation.
This behavior is clearly observed in the frequency spectrum, as the fundamental frequency of the solution is approximately half that of the unstable periodic orbit (red vertical line in Figure \ref{fig:Spectrum_PeriodOrbits}B bottom).

For parameter values $(I_{ext}^E, \varepsilon) =(12,13)$ the system exhibits a regular periodic fast oscillation modulated by a slower rhythm (Figure \ref{fig:Spectrum_PeriodOrbits}C, top; see also supplementary Figure \ref{fig:PopulationFrequency_13_12}). In this case, the homogeneous periodic solution has a small number of unstable directions (only four), all associated with complex Floquet exponents having distinct imaginary parts. The corresponding power spectrum (Figure \ref{fig:Spectrum_PeriodOrbits}C, bottom) shows the characteristic signature of an amplitude-modulated signal: sidebands appear on both sides of the carrier (fast) frequency, which lies near the frequency of the unstable limit cycle (on the gamma range). The slow modulation occurs at a frequency in the theta range (4-8 Hz), determined by the imaginary part of the Floquet exponents associated with the unstable directions (grey vertical lines). This behavior represents an intriguing emergent pattern: a type of cross-frequency coupling, known as phase-amplitude coupling, commonly observed in brain recordings which has been associated to relevant cognitive functions such as memory and information transmission (see the Discussion for more details). 

For parameter values $(I_{ext}^E, \varepsilon) =(12,12)$ (Figure~\ref{fig:Spectrum_PeriodOrbits}D), the signal exhibits a power spectrum with a similar structure to that in Figure~\ref{fig:Spectrum_PeriodOrbits}C, but with broader spectral peaks, thus signaling a chaotic behavior, also suggested by the corresponding time series (compare Figures~\ref{fig:Spectrum_PeriodOrbits}C and \ref{fig:Spectrum_PeriodOrbits}D, and see also supplementary Figure \ref{fig:PopulationFrequency_12_12}). Interestingly, in the latter case, the unstable periodic orbits presents a considerably higher number unstable directions (71). In both cases 
(Figures~\ref{fig:Spectrum_PeriodOrbits}C-D), the power is higher at twice the fundamental frequency, which can be explained by a secondary peak in the temporal voltage trace (see Figures \ref{fig:PopulationFrequency_13_12}C and \ref{fig:PopulationFrequency_12_12}C).

\section{Discussion}\label{sec:discussion}

In this work, we have analyzed the emergence of complex spatiotemporal patterns in a large-scale brain model. Our approach extends the analysis performed in~\cite{clusella2023complex}, which mostly focused in a model with local dynamics described by a Jansen-Rit neural mass model, commonly used to simulate slower neural rhythms. 
In contrast, here we have considered a next-generation neural mass model with the parameters specifically set to reproduce gamma-band oscillations \cite{reyner2022phase} via the PING mechanism. This choice allows us to focus on the fast excitatory-inhibitory interactions within local circuitry of nodes, and to explore the emergence of slow  amplitude modulation of gamma oscillations, which is thought to facilitate flexible communication between distant brain regions~\cite{Fries15, reyner2022phase}. 

The analysis of the model consisted on two parts. First, we have completely characterized different homogeneous solutions and their stability under uniform perturbations. With the help of \texttt{AUTO-07P} \cite{doedel2007auto}, we unveiled a rich bifurcation diagram with regions of bistability, period-doubling bifurcations, and chaotic behavior. This varied dynamical landscape contrasts with that obtained using the Jansen-Rit model in ~\cite{clusella2023complex}, specially on the emergence of chaos within the homogeneous manifold. 
Then, we rigorously characterized the transverse instabilities of both, homogeneous equilibrium points and periodic orbits, within the large-scale brain model. Interestingly, we found different regions of instabilities emerging from both types of solutions. This, again, differs from previous work, which displayed a single region of transverse
instabilities, and only from periodic orbits \cite{clusella2023complex}.

Additionally, we analyzed the heterogeneous spatiotemporal patterns emerging from the destabilization of the homogeneous states in the full system. The computation of Lyapunov exponents revealed an ubiquitous emergence of high-dimensional chaos, as well as the existence of periodic traveling waves in smaller regions of the parameter space. 

Overall, this work reveals the impact of the modeling choices on the ultimate dynamical behavior of the system, showing that next-generation neural mass models offer a complex and rich bifurcation landscape. Importantly, the present model is intrinsically capable of generating these dynamics without relying on noise \cite{Deco2009, Deco2017}, fixed offset delays \cite{Deco2009, Petkoski2019} or heterogeneities in the nodes \cite{PonceAlvarez2015}.
It also establishes transverse instabilities as an ubiquitous route for the emergence of spatiotemporal dynamics, thus
highlighting the importance of such stability analysis for understanding, controlling, and tuning the behavior of large-scale brain models. 

Several large-scale brain models developed in recent years have primarily sought to reproduce empirical observations by fitting functional connectivity (FC) and structural connectivity (SC) data \cite{Deco2017, Forrester2024}. In such approaches, mathematical analysis often plays a secondary role, since the models’ complexity makes it difficult to identify the mechanisms driving specific dynamical patterns from simulations alone. In contrast, the present work incorporates SC extracted from data but places its main emphasis on mathematical analysis, which provides clearer insight into the principles underlying the observed dynamics and forms the core contribution of this study. The task of fitting the model to empirical FC data is left for future work.

Recent studies have also employed large-scale brain models in which the nodes are governed by a neural mass model similar to the NG-NMM used here \cite{Gerster2021, Perl2023, clusella2023complex,Forrester2024}. In \cite{Forrester2024}, a model with additional gap junctions and synapses is shown to reproduce FC patterns observed in MEG and fMRI, although the analysis is limited to linear stability of homogeneous equilibria near Hopf bifurcations. In \cite{Perl2023}, the emphasis is on the effects of structural and functional heterogeneities, but without a detailed analytical investigation of the underlying dynamics. In \cite{Gerster2021}, the node model is reduced to a single excitatory population, and the analysis focuses on transitions between low- and high-activity states associated with seizure-like events, thereby restricting attention to equilibrium points rather than periodic orbits. 
In contrast, our work extends the analytical treatment to include homogeneous periodic orbits and provides a characterization of emergent complex spatiotemporal patterns through Lyapunov exponents and frequency spectra. Finally, although \cite{clusella2023complex} mostly focuses on classical NMMs, it also briefly reports on transverse instabilities emerging from limit-cycles in a model composed of excitation-inhibition pairs of NG-NMMs to illustrate the generality of the methods.
The present work employs a different system for the single node dynamics, which includes synaptic kinetics, and offers a much detailed and curated analysis of the stability properties of emerging dynamical regimes of the brain model. 

A particularly relevant class of solutions emerging from transverse instabilities of periodic orbits corresponds to nodes showing gamma oscillations whose amplitude are systematically modulated by slower theta (4-8Hz) or alpha (8-12Hz) oscillations (see Figures \ref{fig:Spectrum_PeriodOrbits}C and D). This form of rhythmic activity is commonly observed in the brain and is a type of cross-frequency coupling known as phase-amplitude coupling \cite{buzsaki2006rhythms}. 
Specifically, theta–gamma coupling is believed to play a central role in coordinating neural computation and information processing, particularly within the hippocampus and neocortex \cite{Buzski1996, Lisman2013, SchefferTeixeira2018, Jackson2018}. Remarkably, in our model, this slow modulation emerges spontaneously from interactions among nodes that individually exhibit only intrinsic gamma-generating dynamics, without the need to introduce a separate slower oscillator, as is typically required in other models \cite{Segneri2020,ceni2020}. These findings highlight the crucial role of anatomical connectivity in shaping rhythmic brain patterns and motivate further exploration of its mechanistic implications.

Most long-range projections in the brain, including corticocortical and thalamocortical pathways, are mainly excitatory (glutamatergic) and typically innervate both excitatory and inhibitory neurons in their target regions. This dual targeting plays a critical role in maintaining excitatory/inhibitory (E/I) balance and in shaping overall network activity. In our work, we highlight this connectivity by considering that projections from excitatory populations reach both excitatory and inhibitory neurons across target areas, whereas the previous work \cite{clusella2023complex} focused only on long-range inputs to excitatory populations. 
Future work on this direction could explore the effects of having distinct coupling strengths for the excitatory ($\varepsilon_E$) and inhibitory ($\varepsilon_I$) populations, including the case where the inhibitory population receives no long-range projections at all ($\varepsilon_I=0$). Exploring these variations would help reveal how different connectivity patterns change the dynamical landscape. 

Another interesting direction to explore involves modifying the structural connectivity matrix and examine how the regions of transverse instabilities are affected. This could include comparing changes in these regions when using connectomics data from healthy subjects and those with psychiatric or neurological conditions.

Finally, we remark that we used the same neural mass model to describe the dynamics of each brain region.
This simplification, commonly used in the field, overlooks the intrinsic cellular and dynamical brain heterogeneity.
Nonetheless, fundamental understanding of these simpler brain models and their nontrivial dynamics is
a required step before attempting to analyze the behavior of more sophisticated modeling options.

\section*{Supporting information}

\begin{movie}{Movie1}
Chaotic dynamics in the large-scale brain model~\eqref{eq:NetworkModel} arising from the destabilization of an homogeneous equilibrium point. Spatial representation of the excitatory mean membrane potential of each node $v_{E,j}$ for $j=1,\ldots,N$ from simulations of the model corresponding to the results in Figure~\ref{fig:Spectrum_EqPoints}A.
\end{movie}

\begin{movie}{Movie2}
Traveling wave in the large-scale brain model~\eqref{eq:NetworkModel} arising from the destabilization of an homogeneous equilibrium point. Spatial representation of the excitatory mean membrane potential of each node $v_{E,j}$ for $j=1,\ldots,N$ from simulations of the model corresponding to the results in Figure~\ref{fig:Spectrum_EqPoints}B.
\end{movie}

\begin{movie}{Movie3}
Chaotic dynamics in the large-scale brain model~\eqref{eq:NetworkModel} arising from the destabilization of an homogeneous periodic orbit. Spatial representation of the excitatory mean membrane potential of each node $v_{E,j}$ for $j=1,\ldots,N$ from simulations of the model corresponding to the results in Figure~\ref{fig:Spectrum_PeriodOrbits}A.
\end{movie}

\begin{movie}{Movie4}
Traveling wave in the large-scale brain model~\eqref{eq:NetworkModel} arising from the destabilization of an homogeneous periodic orbit. Spatial representation of the excitatory mean membrane potential of each node $v_{E,j}$ for $j=1,\ldots,N$ from simulations of the model corresponding to the results in Figure~\ref{fig:Spectrum_PeriodOrbits}B.
\end{movie}

\renewcommand{\thefigure}{S\arabic{figure}}
\setcounter{figure}{0} 

\begin{figure}[h!]
    \centering
    \includegraphics[width=1\linewidth]{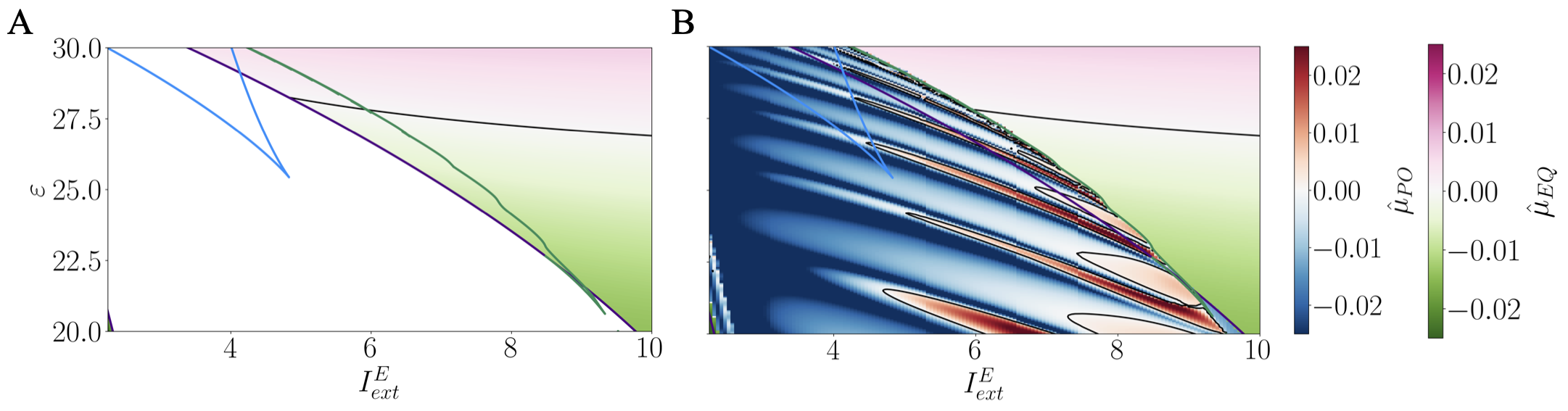}
    \caption{\footnotesize{\textbf{Transverse instabilities in the bistable region.} 
    Largest growth rate, $\hat{\mu}$, for parameter values zoomed around the bistable region. Values for equilibrium points ($\hat{\mu}_{EQ}$) and periodic orbits ($\hat{\mu}_{PO}$) are shown using different color gradients. The bistable region displays the largest growth rate for equilibrium points, $\hat{\mu}_{EQ}$ (\textbf{A}), and for periodic orbits, $\hat{\mu}_{PO}$ (\textbf{B}). In all panels, black curves delimit regions of transverse instabilities, while colored curves denote bifurcations of the homogeneous system, following the same color code as in Figure~\ref{fig:BifDiagram}E.
    }
     }
    \label{fig:Bistability}
\end{figure}

\begin{figure}[h!]
    \centering
    \includegraphics[width=1\linewidth]{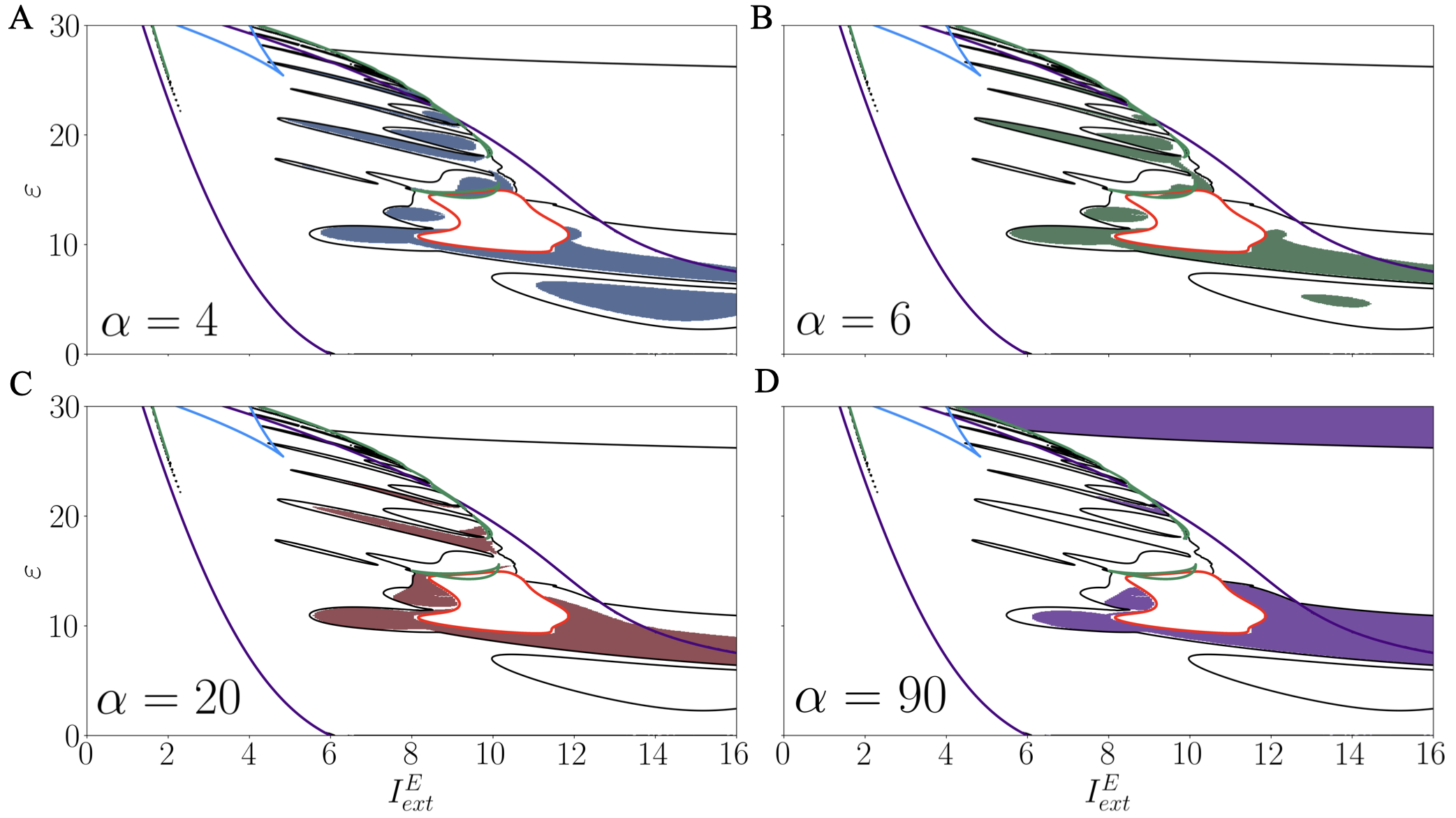}
    \caption{\footnotesize{\textbf{Destabilization of eigenmodes.} Colored regions indicate parameter values where the eigenmode associated with $\Lambda_{4}$ (\textbf{A}), $\Lambda_{6}$ (\textbf{B}), $\Lambda_{20}$ (\textbf{C}), and $\Lambda_{90}$ (\textbf{D}) destabilizes. In all panels, black curves delimit regions of transverse instabilities, while colored curves denote bifurcations of the homogeneous system, following the same color code as in Figure~\ref{fig:BifDiagram}E. 
    }}
    \label{fig:Eigenmodes}
\end{figure}

\begin{figure}[h!]
    \centering
    \includegraphics[width=1\linewidth]{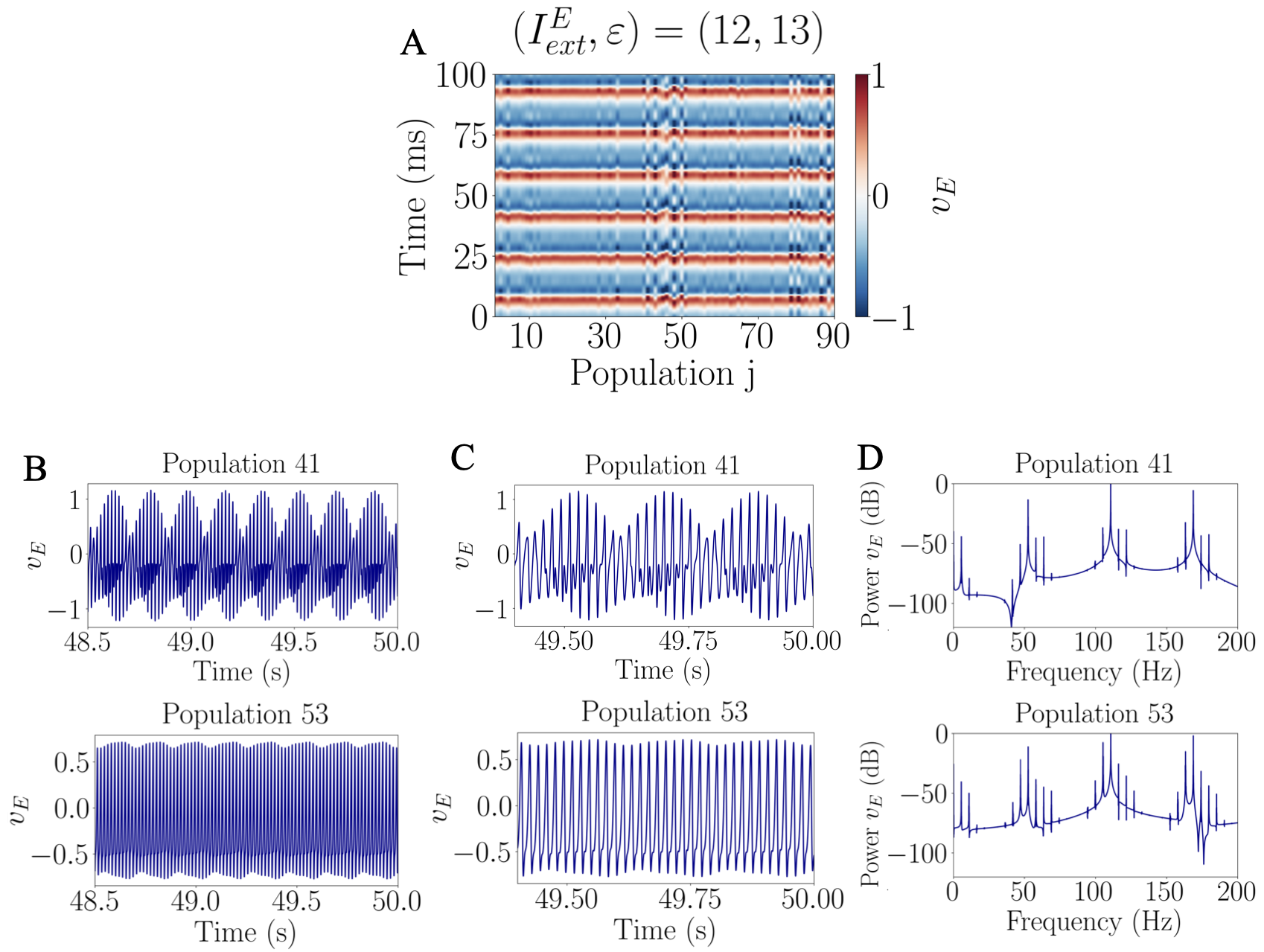}
    \caption{\footnotesize{\textbf{ Activity at individual network nodes for parameters $(I_{ext}^E,\varepsilon)=(12,13)$.} (\textbf{A}) Time series of the excitatory membrane potential at each network node obtained from a simulation of the large-scale brain model~\eqref{eq:NetworkModel} starting near the unstable homogeneous periodic orbit. Colors indicated the magnitude of $v_E$ for a time span of 100 ms. For network nodes 41 (top) and 53 (bottom), (\textbf{B}-\textbf{C}) Times series of the excitatory membrane potential $v_{E,j}$ for a time span of (\textbf{B}) 1.5 s and (\textbf{C}) and 600 ms, (\textbf{D}) Power spectrum of the signal shown in (\textbf{B}) but using a simulation of 50 seconds.}}
    \label{fig:PopulationFrequency_13_12}
\end{figure}

\begin{figure}[h!]
    \centering
    \includegraphics[width=1\linewidth]{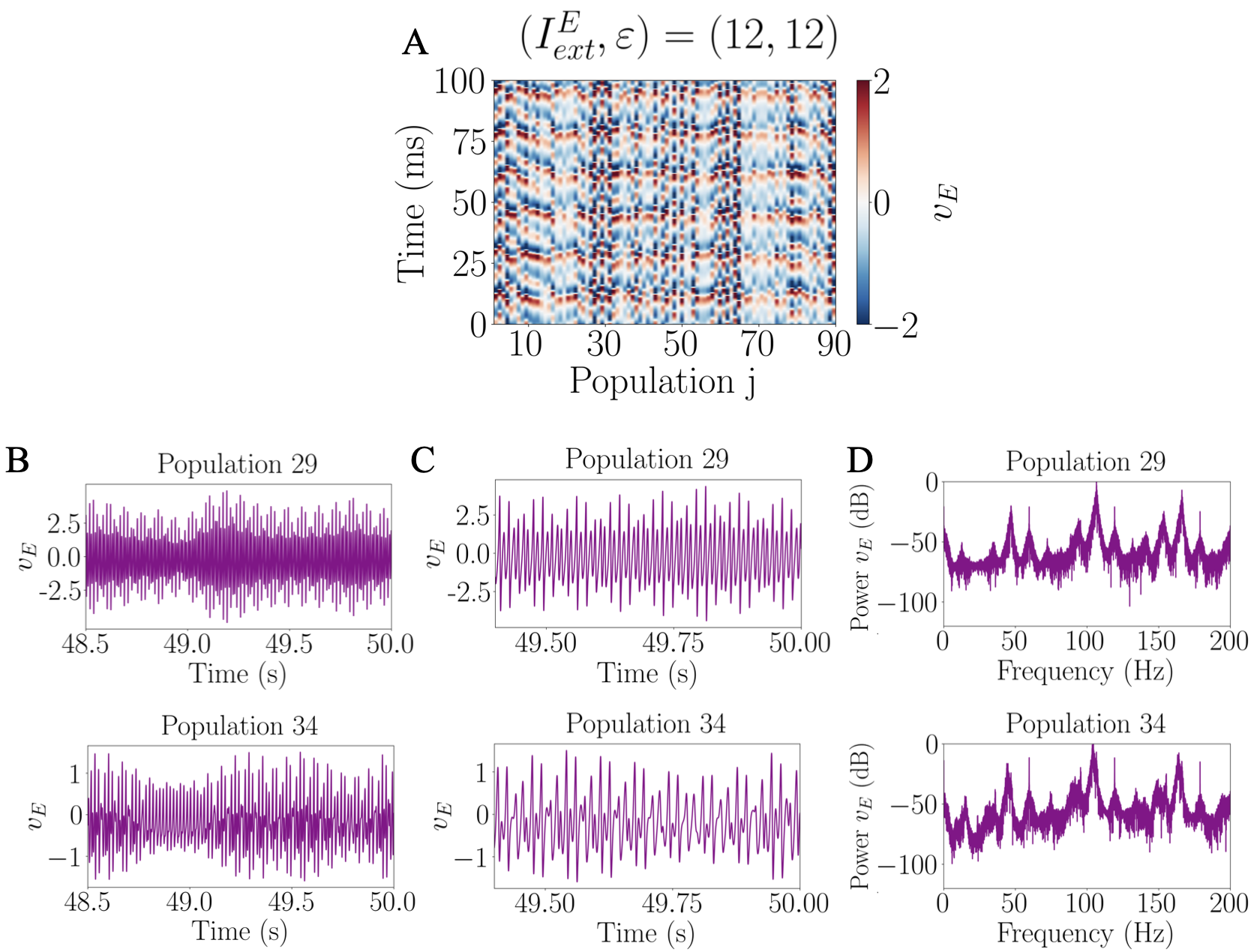}
    \caption{\footnotesize{\textbf{ Activity at individual network nodes for $(I_{ext}^E,\varepsilon)=(12,12)$}. (\textbf{A}) Times series of the excitatory membrane potential at each network node obtained from a simulation of the large-scale brain model~\eqref{eq:NetworkModel} starting near the unstable homogeneous periodic orbit. Colors indicated the magnitude of $v_E$ for a time span of 100 ms. For network nodes 29 (top) and 34 (bottom), (\textbf{B}-\textbf{C}) Times series of the excitatory membrane potential $v_{E,j}$ for a time span of (\textbf{B}) 1.5 s and (\textbf{C}) and 600 ms, (\textbf{D}) Power spectrum of the signal shown in (\textbf{B}) but using a simulation of 50 seconds.}}
    \label{fig:PopulationFrequency_12_12}
\end{figure}

\section*{Acknowledgements}
Work produced with the support of the grants PID2024-155942NB-I00 and PID-2021-122954NB-I00 funded by MCIN/AEI/ 10.13039/501100011033 and “ERDF: A way of making Europe”, the Maria de Maeztu Award for Centers and Units of Excellence in R\&D (CEX2020-001084-M). We also acknowledge the use of the UPC Dynamical Systems group’s cluster for research computing \url{https://dynamicalsystems.upc.edu/en/computing/}.

\appendix
\section*{Appendix} \label{sec:Appendix}

\section{Linear stability analysis}\label{ap:MSF}

Consider $y^{(0)}=(y_0^{(0)},\ldots,y_5^{(0)})^T$ a solution of the homogeneous system~\eqref{eq:HomogeneousSystem} (either an equilibrium point or a periodic orbit) and
\begin{equation}
    \label{eq:SolHomNetwork}
    \bm{y}^{(0)} = (\underbrace{y_0^{(0)},\ldots,y_5^{(0)}}_{\text{ Node $1$ }},\ldots,\underbrace{y_0^{(0)},\ldots,y_5^{(0)}}_{\text{ Node $N$ }})^T,
\end{equation}
a homogeneous solution of the network model~\eqref{eq:NetworkModel}. Consider an arbitrary small perturbation of $\bm{y}^{(0)}$, 
\begin{equation}
    \label{eq:PerturbNetwork}
    \bm{\delta y} = (\underbrace{\delta y_{0,1},\ldots, \delta y_{5,1}}_{\text{ Node $1$ }},\ldots,\underbrace{\delta y_{0,N},\ldots,\delta y_{5,N}}_{\text{ Node $N$ }})^T,
\end{equation}
where each component $\delta y_{k,j}$ represents the perturbation acting on the $k$-th variable of the $j$-th node. Then, the linear evolution of the perturbation is given by
\begin{align}
\label{eq:DynamicsPerturbationVector}
    \frac{d}{dt} \bm{\delta y} (t)=\bm{J}(\bm{y}^{(0)})\bm{\delta y}(t),    
\end{align}
where $\bm{J}$ is the full $6N \times 6N$ Jacobian matrix of the network system~\eqref{eq:NetworkModel} evaluated at $\bm{y}^{(0)}$. The Jacobian matrix of the network can be decomposed in a term containing the Jacobian of the uncoupled system for each node and a coupling term as follows
\begin{equation}
    \label{eq:JacobianDecomposition}
    \bm{J}(\bm{y}^{(0)}) = \bm{I}_N \otimes J(y^{(0)})+\varepsilon \tilde{\bm{W}} \otimes K,
\end{equation}
where $\otimes$ represents the \textit{Kronecker product}, $J$ is the $6 \times 6$ Jacobian matrix of system \eqref{eq:HomogeneousSystem} with $\varepsilon=0$, 
$K=(k_{ij})$ is a $6 \times 6$ matrix whose entries are given by 
\begin{equation}
    k_{ij} = \left\{ \begin{array}{lc} 1 & \text{ if } i \in \{2,5\}, j=3 \\ \\ 0 & \text{ otherwise}, \end{array} \right.
\end{equation}
$\bm{I}_N$ is the $N \times N$ identity matrix, and $\tilde{\bm{W}}$ is the normalized structural connectivity matrix.

In order to simplify the analysis, we follow a well-established method which decouples equation \eqref{eq:DynamicsPerturbationVector} into $N$ first-order variational equations (each of dimension 6), one for each eigenmode of the normalized connectivity matrix $\tilde{\bm{W}}$. To apply the method, the perturbation vector $\bm{\delta y}$ is expressed in the eigenbasis of $\tilde{\bm{W}}$. Let $\bm{\Phi}^{(\alpha)}=(\phi_1^{(\alpha)},\ldots,\phi_N^{(\alpha)})^T$ be a normalized eigenvector with associated eigenvalue $\Lambda_{\alpha}$ for $\alpha = 1,\ldots, N$, then the set of eigenvectors 
$$\bm{P} =\begin{pmatrix}
    \bm{\Phi}^{(1)} |\ldots | \bm{\Phi}^{(j)}|\ldots|\bm{\Phi}^{(N)} 
\end{pmatrix},$$
constitute a basis of the vector space $\mathbb{R}^N$, and we have $\tilde{\bm{W}}\bm{P} = \bm{P}\bm{D}$ with $\bm{D} = \text{diag}(\Lambda_1,\ldots,\Lambda_N)$. Let $u^{(\alpha)}(t) = (u_0^{(\alpha)}(t),\ldots,u_5^{(\alpha)}(t))^T$ for $\alpha=1,\dots,N$ be the coordinates of $\bm{\delta y}(t)$ in this new basis. Then the perturbation of variable $k$ at node $j$ writes as
$$\delta y_{k,j}(t) = \displaystyle \sum_{\alpha=1}^Nu_k^{(\alpha)}(t)\phi_j^{(\alpha)},$$
and the global perturbation vector for all the network becomes
\begin{equation}
\label{eq:NewBasisPerturbationVector}
\bm{\delta y}(t) =\displaystyle \sum_{\alpha=1}^N \bm{\Phi}^{(\alpha)} \otimes u^{(\alpha)}(t).
\end{equation}
Expressing system~\eqref{eq:DynamicsPerturbationVector} in this basis yields the following $N$ variational equations \cite{clusella2023complex}, 
\begin{equation}
\label{eq:MSF_Variationals}
    \frac{d}{dt}u^{(\alpha)}(t) = \mathcal{J}(y^{(0)},\Lambda_{\alpha}) \, u^{(\alpha)}(t), \quad \text{ for } \alpha = 1, \ldots, N, 
\end{equation}
where 
\begin{align}
\mathcal{J}(y^{(0)},\Lambda_{\alpha}) := J(y^{(0)})+\varepsilon \Lambda_{\alpha}K, \nonumber
\end{align}
is a family of $6\times 6$ Jacobians that depend on the homogeneous state of the system $y^{(0)}$ and on the eigenvalues of the structural connectivity matrix 
$\Lambda_{\alpha}$.
If we consider $y^{(0)}$ to be a fixed point, then we can analyze the stability of the homogeneous solution by studying the eigenvalues of the Jacobian $\mathcal{J}$. If we consider $y^{(0)}=y^{(0)}(t)$ to be a periodic solution, then $\mathcal{J}$ is a periodic matrix and Floquet theory applies.

Our procedure consists of computing the largest real part of the eigenvalues of $\mathcal{J}$ (for equilibrium points) or Floquet exponent (for periodic orbits) for each $\alpha=1,\ldots, N$, that we denote $\mu_{max}^{(\alpha)}$ thus obtaining a dispersion relation that characterizes the stability of each mode. 
Therefore we obtain the largest growth rate $\mu_{max}^{(\alpha)}$ of a perturbation acting along each eigenmode $\bm{\Phi}^{(\alpha)}$ as a function of its associated eigenvalue $\Lambda_{\alpha}$ (Eq. \ref{eq:MSF} in the main text). For the case of periodic orbits, this relation
is known as Master Stability Function (MSF) \cite{pecora_master_1998}.

\subsection{Numerical computation of the MSF for periodic orbits}\label{ap:MSF_num}

In the case of periodic orbits we compute the MSF as follows: 
First, we compute an initial condition lying on a periodic orbit $y_0(t)$, as well as its period $T$. 
The variational equations \eqref{eq:MSF_Variationals}, consist of a 6-dimensional system of linear periodic differential equations.
Thus, we compute a fundamental matrix by solving the initial value problem
\[\dfrac{d}{dt} U_{\alpha}(t)=\mathcal{J}(y_0(t),\Lambda_{\alpha}) \, U_{\alpha}(t), \quad U_{\alpha}(0)=I_{6\times6},\]
associated to each eigenvalue $\Lambda_{\alpha}$, for $\alpha = 1, \ldots, N$.  Then, we compute the $6$ eigenvalues of the monodromy matrix $U_{\alpha}(T)$, obtaining the Floquet multipliers $\{\sigma^{(\alpha)}_i\}_{i=1}^6$, and take the largest real part of the corresponding Floquet exponents as 
$$\mu_{max}^{(\alpha)} = \max_{i=1,\ldots,6}\left \{\operatorname{Re}\left(\frac{1}{T}\log{\sigma_i^{(\alpha)}} \right)\right \},$$ 
where $\log$ is the complex logarithm. Repeating this procedure for every $\alpha=1,\ldots, N$, provides us with the Master Stability Function~\eqref{eq:MSF}, each $\alpha$ capturing the largest growth rate along each eigenmode.

\section{Numerical methods}\label{sec:NumericalMethods}

Equations for the homogeneous system~\eqref{eq:HomogeneousSystem} were integrated numerically in Python~\cite{python} using the Runge-Kutta method of order 4–5 (RK45), while the equations of the large-scale brain model~\eqref{eq:NetworkModel} were integrated numerically in Julia~\cite{bezanson2017julia}, which offers improved computational performance, using the Runge-Kutta method of order 4 (RK4) with an absolute tolerance ranging between order $10^{-8}$ and $10^{-9}$. All the simulations of the large-scale brain model have been run for $50$ seconds after discarding an initial transient time of $10$ seconds. The initial conditions have been set by selecting an homogeneous state and adding an independent random perturbation to each variable normally distributed with $0$ mean and $0.01$ standard deviation.

Bifurcation diagrams of the homogeneous system \eqref{eq:HomogeneousSystem} were obtained using the bifurcation analysis software AUTO-07p~\cite{doedel2007auto}. 

Lyapunov exponents were computed using the Julia package ChaosTools~\cite{datseris2018dynamicalsystems,datseris2023chaostools}. Simulations for Lyapunov exponents
of the full network consisted of 4 seconds of computation after discarding a 1 second transient, 
with a time step of $dt = 0.1$ ms. Time windows between consecutive QR-decomposition calls were set to 10 ms. Lyapunov exponents were considered zero if their absolute value fell below the tolerance $10^{-3}$.

All codes used in this work have been released and are publicly available at \url{https://github.com/RosaDelic/LargeScaleBrainModelAnalysis.git}.

\bibliographystyle{abbrv}
\bibliography{references}

\end{document}